\documentclass[12pt,reqno]{article}
\PassOptionsToPackage{linktocpage}{hyperref}
\usepackage[numbers,sort&compress]{natbib}
\usepackage{setspace}
\usepackage{fullpage}
\usepackage{physics}
\usepackage{amsmath}
\usepackage{tikz}
\usepackage{mathdots}
\usepackage{yhmath}
\usepackage{cancel}
\usepackage{color}
\usepackage{siunitx}
\usepackage{array}
\usepackage{multirow}
\usepackage{caption}
\usepackage{amssymb}
\usepackage{array}
\usepackage{textcomp,gensymb}
\usepackage{makecell}
\usepackage{makecell}

\usepackage{float}
\usepackage{cancel}
\usepackage{longtable}
\usepackage{tabularx}
\usepackage{booktabs}
\usetikzlibrary{fadings}
\usepackage{graphicx}
\usepackage{float}
\usepackage[hidelinks]{hyperref}
\usepackage{comment}
\usepackage{physics}
\usepackage{amsmath}
\usepackage{tikz}
\usepackage{mathdots}
\usepackage{yhmath}
\usepackage{cancel}
\usepackage{color}
\usepackage{siunitx}
\usepackage{array}
\usepackage{epsfig}
\usepackage{multirow}
\usepackage{amssymb}
\usepackage{gensymb}
\usepackage{booktabs}
\usetikzlibrary{fadings}
\usepackage{tensor}
\usepackage{ltablex}
\usetikzlibrary{patterns}
\usetikzlibrary{shadows.blur}
\usetikzlibrary{shapes}
\usepackage{appendix}
\usepackage{mdframed}
\mdfdefinestyle{statementstyle}{
   frametitlerule = true,
   frametitlefont = {\normalfont\bfseries},
   frametitlebackgroundcolor = blue!10,
}

\mdtheorem[style=statementstyle]{statement}{WGC as separation of energy scales in quantum gravity}
\mdtheorem[style=statementstyle]{statement2}{Connection between Swampland conjectures from vacuum fluctuations}
\mdtheorem[style=statementstyle]{statement3}{Classical Dominance}
\begin{document}

\begin{titlepage}
\flushright RCHEP-23-001\\
\hfill \\
\vspace*{15mm}
\begin{center}
{\Large \bf Locality Outside Extremal Black Holes}

\vspace*{15mm}

{\large $\text{Hessamaddin Arfaei}^{a,}\footnote{\href{mailto:arfaei@sharif.edu}{arfaei@sharif.edu}}~\text{, Alek Bedroya}^{b,}\footnote{\href{mailto:abedroya@g.harvard.edu}{abedroya@g.harvard.edu}} ~\text{, Mahdi Torabian}^{a,}\footnote{\href{mailto:mahdi@physics.sharif.edu}{mahdi@physics.sharif.edu}}$}
\vspace*{8mm}

\parbox{ \linewidth}{\begin{center}$^a$Research Center for High Energy Physics, Department of Physics, \\Sharif University of Technology, Azadi Ave, Tehran, Iran\\\vspace{0.5cm} $^b$Jefferson Physical Laboratory, Harvard University,\\ Cambridge, MA 02138, USA\end{center}}\\

\vspace*{0.7cm}

\end{center}
\begin{abstract}
Motivated by string theory, we propose that non-local quantum corrections to large extremal black holes must be suppressed by local higher-derivative terms (classical corrections). We show that this condition implies the species bound in all even dimensions, is motivated by Weak Gravity Conjecture, and is necessary for the mild form of the Weak Gravity Conjecture in supersymmetric theories with more than $8$ supercharges. 
\end{abstract}

\end{titlepage}

\tableofcontents

\section{Introduction}\label{sec1}

The recent progress in the Swampland program provides increasing evidence that low energy field theories that are consistent with quantum gravity inherit some universal properties from their gravitational UV completion. The Swampland program attempts to find and understand these universal properties \cite{Vafa:2005ui,Agmon:2022thq}. The Swampland conditions are motivated by quantum gravity but might have little field theory justification. This feature allows the Swampland conditions to rule out theories that are completely consistent looking as an effective field theory. The conditions try to capture the rich and restrictive structure of quantum gravity that gets obscured in low energies. Among the IR footprints of a UV completion, are the coefficients of higher derivative terms in the effective action. There has been great progress in putting bounds on these coefficients using unitarity and causality \cite{Bellazzini:2015cra,Adams:2006sv,Bellazzini:2019xts,Hamada:2018dde,Arkani-Hamed:2021ajd}; however,  much more structure remains to be uncovered. In all the known effective field theories with gravitational UV completion these coefficients are highly correlated, suggesting a rigid underlying structure. 

The coefficients of higher derivative terms contain a lot of information about a theory, however, in this work we study a different type of corrections to the effective action; the quantum corrections. These are all the corrections to the equations of motion that cannot be captured by a local term in the action (e.g. logarithmic running of the coupling). Let us make this distinction sharper. Suppose we want to calculate a specific 1PI scattering vertex $\mathcal{M}$ which at tree level has a polynomial momentum dependence $f(p_i^\mu)$. The 1-PI amplitudes determine the effective action which in turn determines the quantum corrected equations of motion. These are the equations one needs to solve to find the expectation values of different operators in the presence of a source. As we will see later, sometimes one can bypass evaluation of the action and directly read-off the equations of motion. The scaling coefficient $a_\mathcal{M}(E)\simeq\mathcal{M}(E)/f(E)$ typically receives quantum corrections from loops and runs as 
\begin{align}
a_\mathcal{M}(E)\simeq a_{r}+a_{\beta}\ln(E/\mu),
\end{align}
where the running coefficient $a_{\beta}$ is computable from field theory loops while all renormalized values $a_{r}$ are needed as input. We usually think of $a_r$ as the renormalized coefficient of the appropriate term in the effective action. However, there are situations where the momentum dependence of the amplitude is such that it does not correspond to any local terms in the effective action. In other words, the full quantum effective action $\Gamma$ is not necessarily local \cite{Schwartz:2014sze}. Since such terms are typically sub-leading, they can be ignored at low enough energies. An important example of a non-local contribution to amplitude is trace anomaly which we will repeatedly use in this paper. Trace anomaly is the radiative correction to the vev $\expval{T^\mu_\mu}$ of the trace of the energy-momentum tensor. In four dimensions, the correction is
\begin{align}\label{TA3}
\delta T_q=&\frac{1}{180(4\pi)^2}(c\cdot W^2-a\cdot\mathcal{E}_4),
\end{align}
where $(a,c)$ are anomaly coefficients, $\mathcal{E}_4$ is the 4-D Euler density $W_{\mu\nu\rho\sigma}$ is the Weyl tensor. The subscript $q$ stands for quantum, since loop contributions are associated with quantum corrections. Since energy-momentum tensor couples to graviton, the same diagrams that give \eqref{TA3} also give correction to the 1PI effective action and the equations of motion. An equivalent way of finding the modified equations of motion is to plug in the above vev for the energy momentum tensor into the Eintein's field equations. 
\begin{align}\label{ceta}
    R=\kappa (T+\delta T_q).
\end{align}
This turns out to be a non-local correction because there is no local term that can be added to the action which corrects the equations of motion as \eqref{ceta}. It is worth noting that there is a way around this obstruction. If we add a dilaton field, we can write the following action which adds \eqref{TA3} to the trace of Einstein field equations \cite{Komargodski:2011vj}.  
\begin{align}
    S_{anomaly}=&-a\int d^4x\sqrt{- g}(
 \tau E_4+ 4(\mathcal{R}^{\mu\nu}-\frac{1}{2} g^{\mu\nu}
\mathcal{R})\partial_\mu \tau\partial_\nu \tau-4(\partial
\tau)^2\square\
 \tau+2(\partial \tau)^4)\cr&+c\int d^4x \sqrt{-g}  \tau
W_{\mu\nu\rho\sigma}^2 ~.
\end{align}

In the absense of the dilaton, the non-local correction is a universal feature. However, it is suppressed by powers of $M_p$ which makes it negligible at small curvatures.

Based on the above explanation, we can schematically write any amplitude with given external momenta $p^\mu_i$ as
\begin{align}\label{decom}
\mathcal{M}(p_i^\mu)\simeq \sum_k (a_{local}^k+a^k_{\beta}\ln(E/\Lambda_{IR}))f^k_{local}(p_i^\mu)+a^k_{non-local}f^k_{non-local}(p_i^\mu),
\end{align}
where $f^k\sim E^k$ and $a_{local}$ represent the contribution of renormalized coefficients of local higher derivative gravitational terms in the effective action in some IR scale. Note that $a_{non-local}$ and $a_\beta$ are radiative corrections that we can estimate in field theory while $a_{local}$ must be determined by the underlying theory of quantum gravity. We call $a_{local}$ the classical contribution while $a_{\beta}$ and $a_{non-local}$ that come from loops are quantum corrections.

Now that we have defined what we mean by classical and quantum corrections, we are ready to state the main claim of this work. 
\vspace{5pt}

\begin{statement3*}
In the background of large extremal black holes, the non-local corrections to the effective gravitational equations of motion are much smaller than the local corrections of the same order in the mass of the black hole.\vspace{-20pt}\begin{equation}\label{CDS}\end{equation}
\end{statement3*}

The reason we restrict ourselves to extremal black holes is that they have zero Hawking radiation. In fact, the naive generalization of the Classical Dominance to all black holes would have counterexamples with Schwarzchild black holes in four dimensions. However, the extremal black holes in those theories do not violate the Classical Dominance. Effective equations of motion determine the expectation values in presence of a source. Black hole is a gravitational source, therefore, we expect the effective equations of motion to give us the expectation values of operators outside the horizon of a large black hole. 

As we will explain in section \ref{sec2}, the corrections to the equation of motion can be directly calculated in terms of QFT amplitudes that can be decomposed as \eqref{decom}. The classical dominance implies that $|a^k_{local}|\gg|a_{non-local}|,|a_{\beta}|$. If $|a_{non-local}|\gg |a_{local}|$, we can neglect $a_{local}$, and if $|a_{\beta}|\gg |a_{local}|$, we can neglect $a_{local}$ after changing the energy scale by $\mathcal{O}(1)$. In both cases, the information of the gravitational UV theory gets washed away by quantum corrections in the IR. We conjecture that this does not happen in quantum gravity. In particular, this implies the backreaction of trace anomaly is overshadowed by higher derivative terms of the same order in action\footnote{In \cite{Han:2004wt}, based on violation of the unitarity, it was suggested a new physics must enter at high energies to suppress the radiative corrections to gravitational amplitudes which is very much in the same spirit as the Classical Dominance. }.

The Classical Dominance is very natural in quantum gravity since the coefficients of higher derivative terms in the effective action represent tree-level amplitudes while the quantum corrections come from loops. In a unifying UV complete theory, the tree-level amplitudes are expected to suppress the loop contributions.

In fact, in string theory, the higher derivative corrections are proportional to $\alpha'\sim l_s^2$ while the scale of the field theory corrections is $\sim l_P^2$. However we have $l_s\gg l_P$. Otherwise, the string coupling would be $g_s\gg 1$ and string perturbation breaks. Therefore, string theory supports the Classical Dominance. 

\begin{figure}[H]
    \centering

\tikzset{every picture/.style={line width=0.75pt}} 

\begin{tikzpicture}[x=0.75pt,y=0.75pt,yscale=-1,xscale=1]

\draw    (281.81,67.29) .. controls (298.76,84.63) and (314.36,86.9) .. (334.36,66.53) ;
\draw    (293.34,77.09) .. controls (302.83,68.8) and (310.29,67.29) .. (321.82,77.09) ;
\draw    (218.76,80.86) .. controls (283.17,74.07) and (268.93,109.52) .. (310.97,108.02) .. controls (353,106.51) and (346.9,70.3) .. (402.5,75.58) ;
\draw    (366.5,28) .. controls (341.5,53) and (338.77,74.07) .. (401.82,66.53) ;
\draw    (217.4,71.81) .. controls (275.03,67.29) and (283.5,58) .. (229.5,39) ;
\draw   (99.63,50.9) .. controls (118.02,50.35) and (133.27,62.1) .. (133.67,77.14) .. controls (134.07,92.17) and (119.49,104.8) .. (101.09,105.35) .. controls (82.69,105.89) and (67.45,94.14) .. (67.04,79.1) .. controls (66.64,64.07) and (81.23,51.44) .. (99.63,50.9) -- cycle ;
\draw  [fill={rgb, 255:red, 0; green, 0; blue, 0 }  ,fill opacity=1 ] (66.55,77.56) .. controls (67.5,77.53) and (68.29,78.34) .. (68.32,79.36) .. controls (68.35,80.38) and (67.6,81.23) .. (66.65,81.26) .. controls (65.7,81.28) and (64.91,80.48) .. (64.88,79.46) .. controls (64.85,78.44) and (65.6,77.59) .. (66.55,77.56) -- cycle ;
\draw   (177.3,75.28) .. controls (176.4,73.86) and (175.53,72.5) .. (174.55,72.52) .. controls (173.56,72.53) and (172.73,73.91) .. (171.87,75.36) .. controls (171,76.81) and (170.18,78.19) .. (169.19,78.21) .. controls (168.21,78.22) and (167.34,76.86) .. (166.44,75.44) .. controls (165.53,74.02) and (164.66,72.66) .. (163.68,72.68) .. controls (162.7,72.69) and (161.87,74.07) .. (161,75.52) .. controls (160.14,76.97) and (159.31,78.35) .. (158.33,78.37) .. controls (157.34,78.38) and (156.48,77.03) .. (155.57,75.6) .. controls (154.67,74.18) and (153.8,72.83) .. (152.82,72.84) .. controls (151.83,72.86) and (151,74.24) .. (150.14,75.69) .. controls (149.27,77.14) and (148.45,78.52) .. (147.46,78.53) .. controls (146.48,78.55) and (145.61,77.19) .. (144.71,75.77) .. controls (143.8,74.34) and (142.93,72.99) .. (141.95,73) .. controls (140.97,73.02) and (140.14,74.4) .. (139.27,75.85) .. controls (138.41,77.3) and (137.58,78.68) .. (136.6,78.69) .. controls (135.61,78.71) and (134.75,77.35) .. (133.84,75.93) .. controls (133.72,75.74) and (133.59,75.54) .. (133.47,75.35) ;
\draw  [fill={rgb, 255:red, 0; green, 0; blue, 0 }  ,fill opacity=1 ] (133.58,73.74) .. controls (134.53,73.71) and (135.32,74.51) .. (135.35,75.53) .. controls (135.37,76.55) and (134.62,77.4) .. (133.68,77.43) .. controls (132.73,77.46) and (131.93,76.65) .. (131.91,75.63) .. controls (131.88,74.61) and (132.63,73.76) .. (133.58,73.74) -- cycle ;
\draw   (65.1,80.14) .. controls (64.2,78.72) and (63.33,77.36) .. (62.35,77.38) .. controls (61.36,77.39) and (60.54,78.77) .. (59.67,80.22) .. controls (58.81,81.67) and (57.98,83.05) .. (56.99,83.07) .. controls (56.01,83.08) and (55.14,81.73) .. (54.24,80.31) .. controls (53.33,78.88) and (52.46,77.53) .. (51.48,77.54) .. controls (50.5,77.56) and (49.67,78.94) .. (48.81,80.39) .. controls (47.94,81.84) and (47.11,83.22) .. (46.13,83.23) .. controls (45.15,83.25) and (44.28,81.89) .. (43.37,80.47) .. controls (42.47,79.05) and (41.6,77.69) .. (40.62,77.71) .. controls (39.63,77.72) and (38.81,79.1) .. (37.94,80.55) .. controls (37.08,82) and (36.25,83.38) .. (35.26,83.4) .. controls (34.28,83.41) and (33.41,82.06) .. (32.51,80.63) .. controls (31.6,79.21) and (30.73,77.85) .. (29.75,77.87) .. controls (28.77,77.88) and (27.94,79.26) .. (27.08,80.71) .. controls (26.21,82.16) and (25.38,83.54) .. (24.4,83.56) .. controls (23.42,83.57) and (22.55,82.22) .. (21.64,80.8) .. controls (21.52,80.6) and (21.4,80.41) .. (21.27,80.22) ;
\draw  [draw opacity=0][fill={rgb, 255:red, 255; green, 255; blue, 255 }  ,fill opacity=1 ] (81.91,36.67) -- (113.17,36.67) -- (113.17,59.41) -- (81.91,59.41) -- cycle ;
\draw   (33.53,28.31) .. controls (33.44,30.06) and (33.37,31.73) .. (34.29,32.49) .. controls (35.22,33.25) and (36.89,32.88) .. (38.64,32.5) .. controls (40.4,32.11) and (42.07,31.74) .. (43,32.5) .. controls (43.92,33.26) and (43.84,34.93) .. (43.76,36.69) .. controls (43.67,38.44) and (43.59,40.11) .. (44.51,40.87) .. controls (45.44,41.63) and (47.11,41.26) .. (48.87,40.87) .. controls (50.62,40.49) and (52.3,40.12) .. (53.22,40.88) .. controls (54.15,41.64) and (54.07,43.31) .. (53.98,45.06) .. controls (53.89,46.82) and (53.81,48.49) .. (54.74,49.25) .. controls (55.66,50.01) and (57.34,49.64) .. (59.09,49.25) .. controls (60.85,48.86) and (62.52,48.5) .. (63.44,49.26) .. controls (64.37,50.01) and (64.29,51.69) .. (64.2,53.44) .. controls (64.12,55.19) and (64.04,56.87) .. (64.96,57.63) .. controls (65.89,58.38) and (67.56,58.02) .. (69.32,57.63) .. controls (71.07,57.24) and (72.74,56.88) .. (73.67,57.63) .. controls (74.59,58.39) and (74.52,60.06) .. (74.43,61.82) .. controls (74.42,62.06) and (74.41,62.3) .. (74.4,62.53) ;
\draw   (168.95,25.28) .. controls (168.95,25.28) and (168.95,25.28) .. (168.95,25.28) .. controls (169.05,27.05) and (169.14,28.74) .. (168.21,29.49) .. controls (167.29,30.25) and (165.6,29.87) .. (163.84,29.47) .. controls (162.07,29.07) and (160.39,28.69) .. (159.46,29.45) .. controls (158.54,30.21) and (158.63,31.89) .. (158.73,33.66) .. controls (158.82,35.43) and (158.91,37.11) .. (157.99,37.87) .. controls (157.06,38.63) and (155.38,38.25) .. (153.61,37.85) .. controls (151.85,37.45) and (150.16,37.07) .. (149.24,37.83) .. controls (148.31,38.58) and (148.4,40.27) .. (148.5,42.04) .. controls (148.6,43.81) and (148.69,45.49) .. (147.76,46.25) .. controls (146.84,47.01) and (145.16,46.63) .. (143.39,46.23) .. controls (141.62,45.83) and (139.94,45.45) .. (139.01,46.2) .. controls (138.09,46.96) and (138.18,48.65) .. (138.28,50.42) .. controls (138.38,52.18) and (138.47,53.87) .. (137.54,54.63) .. controls (136.62,55.38) and (134.93,55.01) .. (133.16,54.6) .. controls (131.4,54.2) and (129.71,53.82) .. (128.79,54.58) .. controls (127.86,55.34) and (127.95,57.03) .. (128.05,58.79) .. controls (128.07,59.04) and (128.08,59.28) .. (128.09,59.51) ;
\draw  [fill={rgb, 255:red, 0; green, 0; blue, 0 }  ,fill opacity=1 ] (125.65,58.74) .. controls (126.6,58.74) and (127.36,59.58) .. (127.35,60.6) .. controls (127.33,61.62) and (126.55,62.44) .. (125.6,62.43) .. controls (124.65,62.42) and (123.89,61.59) .. (123.9,60.57) .. controls (123.92,59.55) and (124.7,58.73) .. (125.65,58.74) -- cycle ;
\draw  [fill={rgb, 255:red, 0; green, 0; blue, 0 }  ,fill opacity=1 ] (74.58,59.74) .. controls (75.53,59.71) and (76.32,60.51) .. (76.35,61.53) .. controls (76.37,62.55) and (75.62,63.4) .. (74.68,63.43) .. controls (73.73,63.46) and (72.93,62.65) .. (72.91,61.63) .. controls (72.88,60.61) and (73.63,59.76) .. (74.58,59.74) -- cycle ;
\draw    (322.5,34) .. controls (335.5,38) and (337.5,49) .. (359.5,25) ;
\draw    (239.5,31) .. controls (288.5,56) and (270.5,43) .. (288.5,36) ;
\draw    (462.76,83.86) .. controls (527.17,77.07) and (512.93,112.52) .. (554.97,111.02) .. controls (597,109.51) and (590.9,73.3) .. (646.5,78.58) ;
\draw    (610.5,31) .. controls (585.5,56) and (582.77,77.07) .. (645.82,69.53) ;
\draw    (461.4,74.81) .. controls (519.03,70.29) and (527.5,61) .. (473.5,42) ;
\draw    (566.5,37) .. controls (579.5,41) and (581.5,52) .. (603.5,28) ;
\draw    (483.5,34) .. controls (532.5,59) and (514.5,46) .. (532.5,39) ;
\draw    (132.54,202.92) -- (166.54,235.92)(130.46,205.08) -- (164.46,238.08) ;
\draw    (537.6,211.02) -- (510.6,240.02)(535.4,208.98) -- (508.4,237.98) ;

\draw (296.85,32.12) node [anchor=north west][inner sep=0.75pt]  [font=\Large] [align=left] {...};
\draw (86.58,49.17) node [anchor=north west][inner sep=0.75pt]  [font=\LARGE] [align=left] {...};
\draw (181.22,59.62) node [anchor=north west][inner sep=0.75pt]  [font=\Large]  {$\subset $};
\draw (415,59.4) node [anchor=north west][inner sep=0.75pt]  [font=\Large]  {$< $};
\draw (541.85,35.12) node [anchor=north west][inner sep=0.75pt]  [font=\Large] [align=left] {...};
\draw (60,129) node [anchor=north west][inner sep=0.75pt]  [color={rgb, 255:red, 40; green, 9; blue, 226 }  ,opacity=1 ] [align=left] {\begin{minipage}[lt]{58.85pt}\setlength\topsep{0pt}
\begin{center}
Field theory \\loop level
\end{center}

\end{minipage}};
\draw (258,129) node [anchor=north west][inner sep=0.75pt]   [align=left] {\begin{minipage}[lt]{80.39pt}\setlength\topsep{0pt}
\begin{center}
Quantum gravity \\loop level
\end{center}

\end{minipage}};
\draw (505,128) node [anchor=north west][inner sep=0.75pt]  [color={rgb, 255:red, 208; green, 2; blue, 27 }  ,opacity=1 ] [align=left] {\begin{minipage}[lt]{80.39pt}\setlength\topsep{0pt}
\begin{center}
Quantum gravity \\tree level
\end{center}

\end{minipage}};
\draw (142,245) node [anchor=north west][inner sep=0.75pt]  [color={rgb, 255:red, 40; green, 9; blue, 226 }  ,opacity=1 ] [align=left] {Quantum fluctuations};
\draw (380,245) node [anchor=north west][inner sep=0.75pt]  [color={rgb, 255:red, 208; green, 2; blue, 27 }  ,opacity=1 ] [align=left] {Higher derivative terms};
\draw (312,245.4) node [anchor=north west][inner sep=0.75pt]    {$< $};

\end{tikzpicture}
    \caption{The backreaction of vacuum fluctuations captures the contribution of matter loops to gravitational amplitudes. It is natural to expect that such contributions come from a subleading (loop) contribution in the underlying quantum gravity and are dominated by the leading (tree-level) contributions of the same derivative order.}
    \label{TA1}
\end{figure}
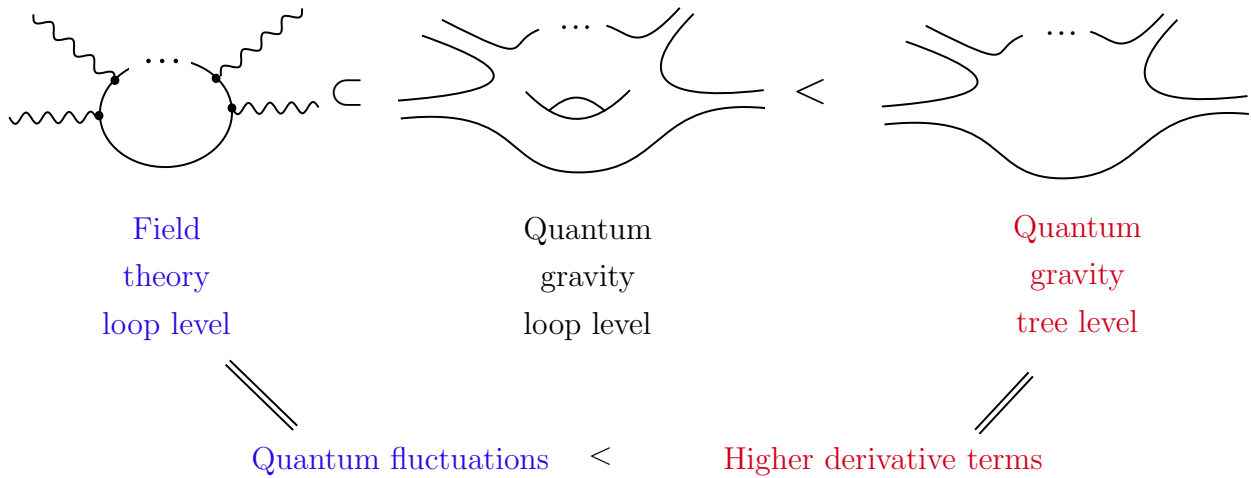

It might seem that our statement is trivial and does not lead to any non-trivial consequence. However, we show that it have various non-trivial implications and is closely related to other Swampland conditions. One of the most notable implications of claim \eqref{CDS} is the species bound \cite{Dvali:2007wp} which in four dimensions states,
\begin{align}
    N_{species}<(\frac{M_P}{\Lambda_{EFT}})^2,
\end{align}
where $N_{species}$ is the number of species and $\Lambda_{EFT}$ is the field theory cutoff. We find that the Classical Dominance implies the species bound and its generalization to all even dimensions with the correct exponents. Given that one of the original derivations of the species bound was based on a black hole entropy formula \cite{Dvali:2007wp}, the Classical Dominance knows about black hole area law which is surprising. 

We also find that some of the implications of the Classical Dominance is required by a version of Weak Gravity Conjecture that states the charge to mass ratio of smooth extremal black holes is bigger than one in Planck units \cite{ArkaniHamed:2006dz,Ma:2021opb,Cano:2019oma,Charles:2016wjs,Aalsma:2021qga,Mirbabayi:2019iae,Aalsma:2020duv}. There have been two types of arguments to verify this version of WGC. The first class of studies test the WGC for effective actions one gets in string theory \cite{Ma:2021opb,Cano:2019oma,Charles:2016wjs}. The second class of studies are bottom-up arguments for the WGC based on general principles such as unitarity. Such arguments aim to show that the higher derivative terms must tilt the extremality curve in the right direction \cite{Arkani-Hamed:2021ajd,Hamada:2018dde,Aalsma:2020duv}. However, in all the arguments we are aware of, the Classical Dominance is implicitly assumed. More precisely, a large scale separation between $a_{local}$ and $\{a_{non-local},a_{\beta}\}$ is assumed which is essentially the statement of the Classical Dominance.

The organization of the paper is as follows. We dedicate the section \ref{sec2} to a general study of different types of corrections to the effective Einstein field equations. We present some analytical methods that allow us to perform a systematic analysis of the implications of different corrections in subsequent sections. In subsections \ref{sec3} we apply the methods of section \ref{sec2} to find the most general correction to extremal black holes. In section \ref{sec5}, we use the results of section \ref{sec3} to formulate and motivate the Classical Dominance and explore its connection to other Swampland conjectures. 

\section{Effective corrections to \texorpdfstring{$\tensor{G}{^\mu_\mu}=\kappa\tensor{T}{^\mu_\mu}$ in 4d}{TEXT}}\label{sec2}

In this section, we study and classify leading order corrections to the Einstein field equations. Among the different corrections that we consider, is quantum corrections to the vacuum expectation value (vev) of energy-momentum tensor. Evaluations of such corrections are often unclear for all components of $T^{\mu\nu}$, but tractable and well-defined for the trace $\expval{\tensor{T}{^\mu^\nu}}g_{\mu\nu}$. For this reason, we focus our attention on corrections to the trace of Einstein field equations. In section \ref{sec3} we will show that for Reissner-Nordstorm backgrounds, 
we can take advantage of the symmetries of spacetime to solve for the corrected spacetime by solely using the corrected trace equation.

There are two types of corrections that can arise for $\tensor{G}{^\mu_\mu}=\kappa\tensor{T}{^\mu_\mu}$; classical and quantum. Classical corrections refer to higher-derivative corrections in the action. Even though these terms are in principle set by the underlying quantum gravity, they function as classical/tree-level correction in the semiclasical field theory. Such corrections trickle down and modify the trace of Einstein field equations. Another type of correction is the backreaction of quantum corrections to the vev of the energy momentum tensor in a fixed background which effectively modifies $\tensor{G}{^\mu_\mu}=\kappa\tensor{T}{_{classical}^\mu_\mu}$. These corrections are associated with radiative corrections to the gravitational amplitudes. 

In this paper we are interested in 
extremal black holes 
. These 
backgrounds have two properties that make them especially suited for our analysis: (i) They have a large symmetry group which allows us to extract a lot of information from the corrected trace equation. (ii) The only physical length scale other than $l_P$ is the horizon's radius, which enables us to do a controlled perturbative study of the corrections \footnote{We work in units where $4\pi\epsilon=\hbar=c=1$. $m_P$ is the Planck mass, while $M_P=m_P/\sqrt{8\pi}$ is the reduced Planck mass.} Note that for non-extremal charged black holes the length scales associated to charge and mass of the black hole are different and not necessarily of the same order. In the rest of this section, we use the notation $L^{-2}$ to refer to the corresponding curvature scale. We will use $l_P/L\ll 1$ as our perturbation parameter.

Suppose the overall correction to the trace component of the Einstein's equations is
\begin{align}
    \tensor{G}{^\mu_\mu}=\kappa(\tensor{T}{^\mu_\mu}+\delta T(\mathcal{R}_{\alpha\beta\rho\sigma},F_{\rho\sigma})),
\end{align}
where $F_{\mu\nu}$ is the field tensor for some $U(1)$ gauge potential and $\delta T$ represents the collective effect of all the corrections to the trace equation. The following discussion of corrections applies to the near horizon geometry of extremal black holes.

We can expand the right side of the above equation in powers of $l_P/L$.
\begin{align}\label{CEOM1}
    \tensor{G}{^\mu_\mu}=\kappa\tensor{T}{^\mu_\mu}+\delta T_0(\mathcal{R}_{\alpha\beta\rho\sigma},F_{\rho\sigma})+\delta T_1(\mathcal{R}_{\alpha\beta\rho\sigma},F_{\rho\sigma})+\delta T_2(\mathcal{R}_{\alpha\beta\rho\sigma},F_{\rho\sigma}))+...
\end{align}
where $\delta T_n=\mathcal{O}((l_P/L)^{2n})$. The reason we only get even powers of $L$ is that $\mathcal{R}_{\alpha\beta\rho\sigma}\sim L^{-2}$ and $F_{\rho\sigma}\sim L^{-1}$. Moreover, we always need even number of field tensors. Thus, we always get even powers of $l_P/L$. 

For $\delta T_0$ to have the right scaling with $L$, it cannot depend on Riemann tensor or the field tensor. Thus, $\delta T_0$ must be a constant scalar. Moreover, it is the trace of some symmetric and divergenceless correction to the Einstein field equations. These constraints fix that tensor to be $\delta T_0^{\mu\nu}=g^{\mu\nu}\delta\Lambda$ for some constant $\delta\Lambda$. Therefore, this term just renormalizes the effective Cosmological constant. We can absorb $\delta\Lambda$ in the definition of $V_{eff}$ the scalar potential. 

Let us take a look at the next order correction, $\delta T_1$. This term is proportional to $L^{-2}$, like $\mathcal{R}$. In fact, the only nonvanishing available scalar quantities of that order are $\mathcal{R}$, $F^2$ \footnote{From the unperturbed Maxwell's equations we know that any other term with the right mass dimension such as $\partial^2F$ or $F\tilde{F}$ vanishes.}. 
\begin{align}\label{fop}
    \delta T_1=A\mathcal{R}+BF^2.
\end{align}
The $\mathcal{R}$ term must be the trace of some symmetric divergenceless correction to the Einstein field equations. According to the Lovelock theorem, these criteria fix that tensor to be a multiple of the Einstein tensor \cite{Lovelock:1971yv,Lovelock:1972vz}. We can absorb such a correction in the renormalization of the Newton's constant $\kappa$. Supposed $\kappa$ is the renormalized Gravitational constant, we can neglect this term. Similarly, we can absorb the second term in the renormalization of the gauge coupling. Thus, if we work with the renormalized energy momentum tensor, the vacuum fluctuations are of subleading order.

\begin{align}
  \tensor{G}{^\mu_\mu}=\kappa(\tensor{T}{^\mu_\mu}+ \delta T_2(\mathcal{R}_{\alpha\beta\rho\sigma},F_{\rho\sigma})).
\end{align}

In the following, we investigate the generic form of $\delta T_2$. We split $\delta T_2$ into a classical part and a quantum part according to our definition in the beginning of the section. 
\begin{align}
    \delta T_2=\delta T_q+\delta T_c,
\end{align}
where, $\delta T_q$ is the quantum correction of order $\mathcal{O}(L^{-2})$ to $\expval{\tensor{T}{^\mu_\mu}}$, and $\delta T_c$ is corrections of order $\mathcal{O}(L^{-2})$ from subleading terms in the classical action. Let us start with $\delta T_q$\footnote{One might worry about the reliance of our definition for classical corrections on the existence of an action. For theories without action, classical corrections refer to the tree-level amplitudes. For example, in string theories, the classical corrections refers to $\alpha'$ corrctions while quantum corrections involve subleading $g_s$ corrections.}.

\begin{figure}
    \centering

\tikzset{every picture/.style={line width=0.75pt}} 

\begin{tikzpicture}[x=0.75pt,y=0.75pt,yscale=-1,xscale=1]

\draw   (69.98,142.38) .. controls (69.98,120.31) and (83.92,102.41) .. (101.12,102.41) .. controls (118.32,102.41) and (132.26,120.31) .. (132.26,142.38) .. controls (132.26,164.45) and (118.32,182.35) .. (101.12,182.35) .. controls (83.92,182.35) and (69.98,164.45) .. (69.98,142.38) -- cycle ;
\draw  [fill={rgb, 255:red, 208; green, 2; blue, 27 }  ,fill opacity=1 ] (99.34,182.89) .. controls (99.34,181.75) and (100.29,180.83) .. (101.45,180.83) .. controls (102.62,180.83) and (103.57,181.75) .. (103.57,182.89) .. controls (103.57,184.03) and (102.62,184.95) .. (101.45,184.95) .. controls (100.29,184.95) and (99.34,184.03) .. (99.34,182.89) -- cycle ;
\draw   (100.46,50.04) .. controls (98.81,51.08) and (97.23,52.08) .. (97.21,53.26) .. controls (97.2,54.44) and (98.75,55.47) .. (100.37,56.56) .. controls (102,57.64) and (103.55,58.68) .. (103.54,59.86) .. controls (103.52,61.04) and (101.94,62.03) .. (100.28,63.07) .. controls (98.63,64.11) and (97.05,65.11) .. (97.03,66.29) .. controls (97.02,67.47) and (98.57,68.51) .. (100.19,69.59) .. controls (101.82,70.67) and (103.37,71.71) .. (103.36,72.89) .. controls (103.34,74.07) and (101.76,75.06) .. (100.1,76.1) .. controls (98.45,77.15) and (96.87,78.14) .. (96.85,79.32) .. controls (96.84,80.5) and (98.39,81.54) .. (100.01,82.62) .. controls (101.64,83.7) and (103.19,84.74) .. (103.18,85.92) .. controls (103.16,87.1) and (101.58,88.1) .. (99.92,89.14) .. controls (98.27,90.18) and (96.69,91.17) .. (96.67,92.35) .. controls (96.66,93.53) and (98.21,94.57) .. (99.83,95.65) .. controls (101.46,96.74) and (103.01,97.77) .. (103,98.95) .. controls (102.98,100.13) and (101.4,101.13) .. (99.75,102.17) .. controls (99.52,102.31) and (99.29,102.45) .. (99.07,102.59) ;
\draw  [fill={rgb, 255:red, 0; green, 0; blue, 0 }  ,fill opacity=1 ] (97.23,102.41) .. controls (97.23,101.28) and (98.17,100.35) .. (99.34,100.35) .. controls (100.51,100.35) and (101.45,101.28) .. (101.45,102.41) .. controls (101.45,103.55) and (100.51,104.48) .. (99.34,104.48) .. controls (98.17,104.48) and (97.23,103.55) .. (97.23,102.41) -- cycle ;
\draw   (217.11,136.51) .. controls (217.11,114.44) and (231.05,96.55) .. (248.25,96.55) .. controls (265.45,96.55) and (279.4,114.44) .. (279.4,136.51) .. controls (279.4,158.59) and (265.45,176.48) .. (248.25,176.48) .. controls (231.05,176.48) and (217.11,158.59) .. (217.11,136.51) -- cycle ;
\draw  [fill={rgb, 255:red, 208; green, 2; blue, 27 }  ,fill opacity=1 ] (246.14,177.02) .. controls (246.14,175.88) and (247.09,174.96) .. (248.25,174.96) .. controls (249.42,174.96) and (250.37,175.88) .. (250.37,177.02) .. controls (250.37,178.16) and (249.42,179.09) .. (248.25,179.09) .. controls (247.09,179.09) and (246.14,178.16) .. (246.14,177.02) -- cycle ;
\draw   (204.53,63.31) .. controls (204.44,65.06) and (204.37,66.73) .. (205.29,67.49) .. controls (206.22,68.25) and (207.89,67.88) .. (209.64,67.5) .. controls (211.4,67.11) and (213.07,66.74) .. (214,67.5) .. controls (214.92,68.26) and (214.84,69.93) .. (214.76,71.69) .. controls (214.67,73.44) and (214.59,75.11) .. (215.51,75.87) .. controls (216.44,76.63) and (218.11,76.26) .. (219.87,75.87) .. controls (221.62,75.49) and (223.3,75.12) .. (224.22,75.88) .. controls (225.15,76.64) and (225.07,78.31) .. (224.98,80.06) .. controls (224.89,81.82) and (224.81,83.49) .. (225.74,84.25) .. controls (226.66,85.01) and (228.34,84.64) .. (230.09,84.25) .. controls (231.85,83.86) and (233.52,83.5) .. (234.44,84.26) .. controls (235.37,85.01) and (235.29,86.69) .. (235.2,88.44) .. controls (235.12,90.19) and (235.04,91.87) .. (235.96,92.63) .. controls (236.89,93.38) and (238.56,93.02) .. (240.32,92.63) .. controls (242.07,92.24) and (243.74,91.88) .. (244.67,92.63) .. controls (245.59,93.39) and (245.52,95.06) .. (245.43,96.82) .. controls (245.42,97.06) and (245.41,97.3) .. (245.4,97.53) ;
\draw  [fill={rgb, 255:red, 0; green, 0; blue, 0 }  ,fill opacity=1 ] (244.03,96.55) .. controls (244.03,95.41) and (244.97,94.48) .. (246.14,94.48) .. controls (247.31,94.48) and (248.25,95.41) .. (248.25,96.55) .. controls (248.25,97.69) and (247.31,98.61) .. (246.14,98.61) .. controls (244.97,98.61) and (244.03,97.69) .. (244.03,96.55) -- cycle ;
\draw   (289.95,63.28) .. controls (290.05,65.05) and (290.14,66.74) .. (289.21,67.49) .. controls (288.29,68.25) and (286.6,67.87) .. (284.84,67.47) .. controls (283.07,67.07) and (281.39,66.69) .. (280.46,67.45) .. controls (279.54,68.21) and (279.63,69.89) .. (279.73,71.66) .. controls (279.82,73.43) and (279.91,75.11) .. (278.99,75.87) .. controls (278.06,76.63) and (276.38,76.25) .. (274.61,75.85) .. controls (272.85,75.45) and (271.16,75.07) .. (270.24,75.83) .. controls (269.31,76.58) and (269.4,78.27) .. (269.5,80.04) .. controls (269.6,81.81) and (269.69,83.49) .. (268.76,84.25) .. controls (267.84,85.01) and (266.16,84.63) .. (264.39,84.23) .. controls (262.62,83.83) and (260.94,83.45) .. (260.01,84.2) .. controls (259.09,84.96) and (259.18,86.65) .. (259.28,88.42) .. controls (259.38,90.18) and (259.47,91.87) .. (258.54,92.63) .. controls (257.62,93.38) and (255.93,93.01) .. (254.16,92.6) .. controls (252.4,92.2) and (250.71,91.82) .. (249.79,92.58) .. controls (248.86,93.34) and (248.95,95.03) .. (249.05,96.79) .. controls (249.07,97.04) and (249.08,97.28) .. (249.09,97.51) ;
\draw   (341.35,137.38) .. controls (341.35,115.31) and (355.3,97.41) .. (372.5,97.41) .. controls (389.7,97.41) and (403.64,115.31) .. (403.64,137.38) .. controls (403.64,159.45) and (389.7,177.35) .. (372.5,177.35) .. controls (355.3,177.35) and (341.35,159.45) .. (341.35,137.38) -- cycle ;
\draw  [fill={rgb, 255:red, 208; green, 2; blue, 27 }  ,fill opacity=1 ] (370.38,177.35) .. controls (370.38,176.21) and (371.33,175.28) .. (372.5,175.28) .. controls (373.66,175.28) and (374.61,176.21) .. (374.61,177.35) .. controls (374.61,178.49) and (373.66,179.41) .. (372.5,179.41) .. controls (371.33,179.41) and (370.38,178.49) .. (370.38,177.35) -- cycle ;
\draw  [fill={rgb, 255:red, 0; green, 0; blue, 0 }  ,fill opacity=1 ] (390.85,108.71) .. controls (390.85,107.57) and (391.8,106.65) .. (392.96,106.65) .. controls (394.13,106.65) and (395.08,107.57) .. (395.08,108.71) .. controls (395.08,109.85) and (394.13,110.77) .. (392.96,110.77) .. controls (391.8,110.77) and (390.85,109.85) .. (390.85,108.71) -- cycle ;
\draw  [fill={rgb, 255:red, 0; green, 0; blue, 0 }  ,fill opacity=1 ] (349.03,108.71) .. controls (349.03,107.57) and (349.97,106.65) .. (351.14,106.65) .. controls (352.31,106.65) and (353.25,107.57) .. (353.25,108.71) .. controls (353.25,109.85) and (352.31,110.77) .. (351.14,110.77) .. controls (349.97,110.77) and (349.03,109.85) .. (349.03,108.71) -- cycle ;
\draw   (308.64,75.47) .. controls (308.56,77.23) and (308.48,78.9) .. (309.4,79.66) .. controls (310.33,80.41) and (312,80.05) .. (313.75,79.66) .. controls (315.51,79.27) and (317.18,78.91) .. (318.11,79.66) .. controls (319.03,80.42) and (318.95,82.1) .. (318.87,83.85) .. controls (318.78,85.6) and (318.7,87.28) .. (319.63,88.03) .. controls (320.55,88.79) and (322.23,88.43) .. (323.98,88.04) .. controls (325.73,87.65) and (327.41,87.28) .. (328.33,88.04) .. controls (329.26,88.8) and (329.18,90.47) .. (329.09,92.23) .. controls (329,93.98) and (328.93,95.65) .. (329.85,96.41) .. controls (330.78,97.17) and (332.45,96.8) .. (334.2,96.42) .. controls (335.96,96.03) and (337.63,95.66) .. (338.56,96.42) .. controls (339.48,97.18) and (339.4,98.85) .. (339.32,100.6) .. controls (339.23,102.36) and (339.15,104.03) .. (340.08,104.79) .. controls (341,105.55) and (342.67,105.18) .. (344.43,104.79) .. controls (346.18,104.41) and (347.86,104.04) .. (348.78,104.8) .. controls (349.71,105.56) and (349.63,107.23) .. (349.54,108.98) .. controls (349.53,109.22) and (349.52,109.46) .. (349.51,109.7) ;
\draw   (434.11,74.58) .. controls (434.2,76.35) and (434.29,78.03) .. (433.37,78.79) .. controls (432.44,79.55) and (430.76,79.17) .. (428.99,78.77) .. controls (427.23,78.37) and (425.54,77.99) .. (424.62,78.74) .. controls (423.69,79.5) and (423.78,81.19) .. (423.88,82.96) .. controls (423.98,84.72) and (424.07,86.41) .. (423.14,87.17) .. controls (422.22,87.92) and (420.53,87.55) .. (418.77,87.14) .. controls (417,86.74) and (415.32,86.36) .. (414.39,87.12) .. controls (413.47,87.88) and (413.56,89.57) .. (413.66,91.33) .. controls (413.75,93.1) and (413.84,94.79) .. (412.92,95.54) .. controls (411.99,96.3) and (410.31,95.92) .. (408.54,95.52) .. controls (406.78,95.12) and (405.09,94.74) .. (404.17,95.5) .. controls (403.24,96.26) and (403.33,97.94) .. (403.43,99.71) .. controls (403.53,101.48) and (403.62,103.16) .. (402.7,103.92) .. controls (401.77,104.68) and (400.09,104.3) .. (398.32,103.9) .. controls (396.55,103.5) and (394.87,103.12) .. (393.94,103.88) .. controls (393.02,104.63) and (393.11,106.32) .. (393.21,108.09) .. controls (393.22,108.33) and (393.23,108.57) .. (393.24,108.81) ;
\draw   (467.71,144.33) .. controls (467.71,122.26) and (481.65,104.37) .. (498.85,104.37) .. controls (516.06,104.37) and (530,122.26) .. (530,144.33) .. controls (530,166.4) and (516.06,184.3) .. (498.85,184.3) .. controls (481.65,184.3) and (467.71,166.4) .. (467.71,144.33) -- cycle ;
\draw  [fill={rgb, 255:red, 208; green, 2; blue, 27 }  ,fill opacity=1 ] (497.08,184.84) .. controls (497.08,183.7) and (498.02,182.78) .. (499.19,182.78) .. controls (500.36,182.78) and (501.3,183.7) .. (501.3,184.84) .. controls (501.3,185.98) and (500.36,186.9) .. (499.19,186.9) .. controls (498.02,186.9) and (497.08,185.98) .. (497.08,184.84) -- cycle ;
\draw   (498.2,51.99) .. controls (496.54,53.03) and (494.96,54.03) .. (494.95,55.21) .. controls (494.93,56.39) and (496.48,57.42) .. (498.11,58.51) .. controls (499.74,59.59) and (501.29,60.63) .. (501.27,61.81) .. controls (501.25,62.99) and (499.68,63.98) .. (498.02,65.02) .. controls (496.36,66.07) and (494.78,67.06) .. (494.77,68.24) .. controls (494.75,69.42) and (496.3,70.46) .. (497.93,71.54) .. controls (499.56,72.62) and (501.11,73.66) .. (501.09,74.84) .. controls (501.07,76.02) and (499.5,77.01) .. (497.84,78.06) .. controls (496.18,79.1) and (494.6,80.09) .. (494.59,81.27) .. controls (494.57,82.45) and (496.12,83.49) .. (497.75,84.57) .. controls (499.38,85.65) and (500.93,86.69) .. (500.91,87.87) .. controls (500.89,89.05) and (499.32,90.05) .. (497.66,91.09) .. controls (496,92.13) and (494.42,93.12) .. (494.41,94.3) .. controls (494.39,95.48) and (495.94,96.52) .. (497.57,97.6) .. controls (499.2,98.69) and (500.75,99.72) .. (500.73,100.9) .. controls (500.71,102.08) and (499.14,103.08) .. (497.48,104.12) .. controls (497.25,104.26) and (497.03,104.4) .. (496.81,104.54) ;
\draw  [fill={rgb, 255:red, 0; green, 0; blue, 0 }  ,fill opacity=1 ] (494.96,104.37) .. controls (494.96,103.23) and (495.91,102.3) .. (497.08,102.3) .. controls (498.24,102.3) and (499.19,103.23) .. (499.19,104.37) .. controls (499.19,105.51) and (498.24,106.43) .. (497.08,106.43) .. controls (495.91,106.43) and (494.96,105.51) .. (494.96,104.37) -- cycle ;
\draw   (499.98,186.66) .. controls (498.32,187.7) and (496.74,188.7) .. (496.73,189.88) .. controls (496.71,191.06) and (498.26,192.09) .. (499.89,193.18) .. controls (501.52,194.26) and (503.07,195.3) .. (503.05,196.48) .. controls (503.03,197.66) and (501.46,198.65) .. (499.8,199.69) .. controls (498.14,200.73) and (496.56,201.73) .. (496.55,202.91) .. controls (496.53,204.09) and (498.08,205.12) .. (499.71,206.21) .. controls (501.34,207.29) and (502.89,208.33) .. (502.87,209.51) .. controls (502.85,210.69) and (501.28,211.68) .. (499.62,212.72) .. controls (497.96,213.77) and (496.38,214.76) .. (496.37,215.94) .. controls (496.35,217.12) and (497.9,218.16) .. (499.53,219.24) .. controls (501.16,220.32) and (502.71,221.36) .. (502.69,222.54) .. controls (502.67,223.72) and (501.1,224.71) .. (499.44,225.76) .. controls (497.78,226.8) and (496.2,227.79) .. (496.19,228.97) .. controls (496.17,230.15) and (497.72,231.19) .. (499.35,232.27) .. controls (500.98,233.36) and (502.53,234.39) .. (502.51,235.57) .. controls (502.49,236.75) and (500.92,237.75) .. (499.26,238.79) .. controls (499.03,238.93) and (498.81,239.07) .. (498.59,239.21) ;

\draw (88.52,188.52) node [anchor=north west][inner sep=0.75pt]    {$T_{( 0) \ \mu \nu }{}$};
\draw (-10,129.24) node [anchor=north west][inner sep=0.75pt]    {$< T_{\mu \nu }> =$};
\draw (235.32,183.52) node [anchor=north west][inner sep=0.75pt]    {$T_{( 0) \ \mu \nu }{}$};
\draw (359.9,183.52) node [anchor=north west][inner sep=0.75pt]    {$T_{( 0) \ \mu \nu }{}$};
\draw (450.33,179.17) node [anchor=north west][inner sep=0.75pt]    {$T_{( 1) \ \mu \nu }{}$};
\draw (175.8,113.96) node [anchor=north west][inner sep=0.75pt]  [font=\Huge] [align=left] {{\huge {\fontfamily{pcr}\selectfont \{}}};
\draw (301.26,125.51) node [anchor=north west][inner sep=0.75pt]  [font=\LARGE] [align=left] {+};
\draw (431.18,125.59) node [anchor=north west][inner sep=0.75pt]  [font=\LARGE] [align=left] {+};
\draw (578.87,155.02) node [anchor=north west][inner sep=0.75pt]  [font=\Huge,rotate=-180] [align=left] {{\huge {\fontfamily{pcr}\selectfont \{}}};
\draw (147.26,121.51) node [anchor=north west][inner sep=0.75pt]  [font=\LARGE] [align=left] {+};
\draw (593.18,126.59) node [anchor=north west][inner sep=0.75pt]  [font=\LARGE] [align=left] {+ ...};

\end{tikzpicture}
    \caption{Diagramatic expansion of $\expval{T_{\mu\nu}}$ for a free theory in a fixed gravitational background. $T_{(i)~\mu\nu}$ is the $i$-graviton part of $T_{\mu\nu}$ in curved spacetime.}
    \label{TA2}
\end{figure}
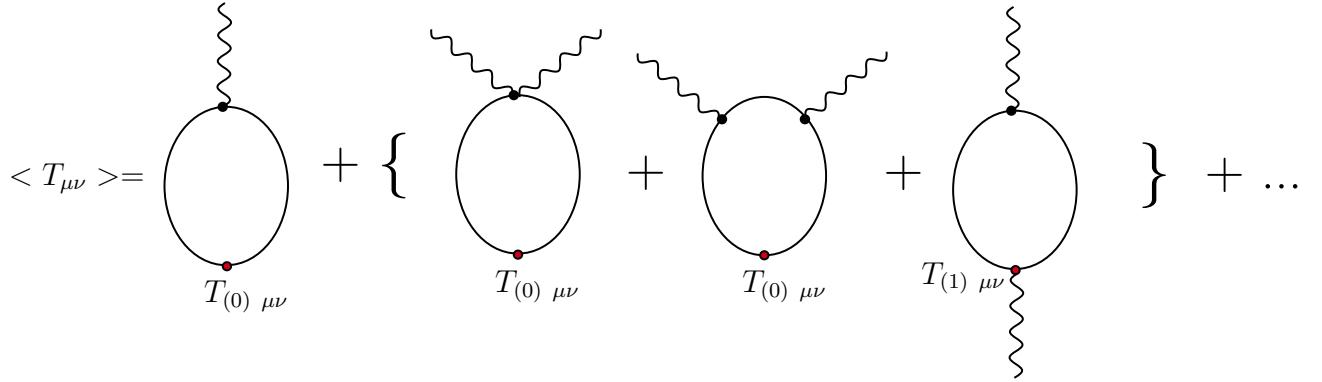

Before proceeding with more detailed study of the corrections, let us clarify the connection between backreaction of vacuum fluctuations and corrections to gravitational amplitudes. As one can see in Figure \ref{TA2}, by attaching an external graviton to diagrams corresponding to $\expval{T_{\mu\nu}}$, we find field theory corrections to gravitational amplitudes. In fact, plugging $\expval{T_{\mu\nu}}$ back into the Einstein equations effectively modifies the dynamics the same way as integrating out matter loops in gravitational vertices would. Thus, considering backreaction of the vacuum fluctuations is equivalent to correcting gravitational amplitudes with matter loops.

\subsection{Quantum corrections}\label{22}

Vacuum expectation value of the energy-momentum tensor $\expval{T^{\mu\nu}}$ is generally UV-divergent and needs to be regularized. The full tensor $\expval{T_{\mu\nu}}$ depends on the regularization scheme. However, when we consider the trace $\expval{\tensor{T}{^\mu_\mu}}$, the divergences can be removed by adding local counterterms to obtain a universal scheme-independent contribution that is called trace anomaly.

Based on dimensional analysis, we can easily see that $\delta T_1$ in equation \eqref{CEOM1} only contains terms quadratic in Riemann curvature tensor or derivatives of the Ricci scalar with coefficients that have mass-dimension zero. A careful calculation of the trace anomaly in $D=4$ gives \cite{Duff:1977ay,Duff:1993wm,Wald:1978pj}, 
\begin{align}
\label{anomalyS}
\delta T_q=&\frac{1}{180(4\pi)^2}(c\cdot W^2-a\cdot\mathcal{E}_4)
\end{align}
where the 4-D Euler density $\mathcal{E}_4$ and the square of the Weyl tensor $W_{\mu\nu\rho\sigma}W^{\mu\nu\rho\sigma}$ are given by,
\begin{align}
W^2&=W_{\mu\nu\rho\sigma}W^{\mu\nu\rho\sigma}=\mathcal{R}_{\mu{}\nu{}\rho{}\sigma{}}\mathcal{R}^{\mu{}\nu{}\rho{}\sigma{}}-2\mathcal{R}_{\mu{}\nu{}}\mathcal{R}^{\mu{}\nu{}}+\frac{1}{3}\mathcal{R}^2,\nonumber\\
\mathcal{E}_4&=\mathcal{R}_{\mu{}\nu{}\rho{}\sigma{}}\mathcal{R}^{\mu{}\nu{}\rho{}\sigma{}}-4\mathcal{R}_{\mu{}\nu{}}\mathcal{R}^{\mu{}\nu{}}+\mathcal{R}^2.
\end{align}

Note that all other terms (e.g. $(F^2)^2$ or $\Box \mathcal{R}$) can be absorbed in renormalization of higher derivative terms \cite{Christensen:1976vb,Adler:1976jx,Wald:1978pj,Wald:1977up,Deser:1976yx,Brown:1976wc,Brown:1977pq,Capper:1974ic,Duff:1977ay,Duff:1993wm,Pascual:1988ri}. Thus, we account for them in the study of classical corrections in subsection \ref{26}. 

As we stated above, the trace anomaly cannot be derived from any local covariant action \cite{Deser:1976yx}, unless we add a dilaton field \cite{Komargodski:2011vj}. With the addition of dilaton $\tau$, one can get the trace anomaly from the following action. 
\begin{align}
    S_{anomaly}=&-a\int d^4x\sqrt{- g}(
 \tau E_4+ 4(\mathcal{R}^{\mu\nu}-\frac{1}{2} g^{\mu\nu}
\mathcal{R})\partial_\mu \tau\partial_\nu \tau-4(\partial
\tau)^2\square\
 \tau+2(\partial \tau)^4)\cr&+c\int d^4x \sqrt{-g}  \tau
W_{\mu\nu\rho\sigma}^2 ~.
\end{align}
This dilaton could be thought of as the scaling degree of freedom of the fields that have been integrated out in the loops\footnote{We are thankful to Cumrun Vafa for explaining this point to us.}. In the presence of dilaton, the anomalies could be absorbed into renormalization of the couplings in the following action. However, in the absence of a dilatonic action such as the one above, that would not be possible. We assume that there is no such light dilatonic field in the field theory. 

One can explicitly verify the above structure for the free field theories \cite{Christensen:1976vb}. For conformal field theories where no mass scale is available, $a$ and $c$ are called the conformal charges since they represent the breakdown of scale invariance at the quantum level. We will adapt this terminology hereinafter even for non-conformal theories. The conformal charges for free theories are given in the Table \ref{table1} \cite{Christensen:1978md,Christensen:1978gi,Meissner:2016onk}.

\begin{center}
\small
  \renewcommand{\arraystretch}{1.5}
    \addtolength{\tabcolsep}{1pt}  \captionsetup{width=.9\linewidth}
\begin{longtable}{|c|c|c|}\hline \hline
\diaghead{xxxxxxxxxxxxxxxxx}{Spin (s)}{Mass (m)} & \makecell{$m=0$} & \makecell{$m\neq0$}\\
\hline\hline
$s=0~\text{(scalar)}$ &$~~~~~(\frac{3}{2},\frac{1}{2})~~~~~$ &$~~~~~(\frac{3}{2},\frac{1}{2})~~~~~$\\\hline
$s=0~\text{(2-form)}$ &$(\frac{3}{2},-\frac{179}{2})$ &$\emptyset$\\\hline
$s=\frac{1}{2}~\text{Majorana}$ &$(\frac{9}{2},\frac{11}{4})$ &$(\frac{9}{2},\frac{11}{4})$\\\hline
$s=\frac{1}{2}~\text{Dirac}$ &$(9,\frac{11}{2})$ &$(9,\frac{11}{2})$\\\hline
$s=1$ &$(18,31)$ &$(\frac{39}{2},\frac{63}{2})$\\\hline
$s=\frac{3}{2} $ &$(-\frac{411}{2},-\frac{589}{4})$ &$(-201,-\frac{289}{2})$\\\hline
$s=2$ &$(783,571)$ &$(\frac{1605}{2},\frac{1205}{2})$\\\hline
 \caption{The conformal charges $(c,a)$ for different fields based on their mass and spin \cite{Meissner:2016onk}.}
\label{table1}
\end{longtable}
\end{center}

\subsection{Classical corrections}\label{26}

In this section we consider higher derivative corrections to the effective action. Such terms naturally arise in string theory as $\alpha'$ corrections. Thus, from the perspective of quantum gravity it is important to take them into account in subleading calculations.

The effective field theory might have various fields. However, we are interested in only two fields; gravity and a $U(1)$ gauge potential. We assume the other fields are set to their vacuum. In a quantum mechanical treatment, we have to take into account the quantum fluctuations of all the fields which is basically what we did in the previous subsection.

We start by finding the correction terms that contribute to the first subleading order. We are interested in terms whose contribution to the equation of motion is $\mathcal{O}((l_P/L)^{-4})$. Following the dimensional analysis in the beginning of the section, we can find that such corrections can only come from a Lagrangian with the following terms \cite{Kats:2006xp},

\begin{align}
\label{GeneralCS}
\delta S\simeq\int d^4x &\sqrt{-g}(c_1\mathcal{R}^2+c_2\mathcal{R}_{\mu\nu}\mathcal{R}^{\mu\nu}+c_3\mathcal{R}_{\mu\nu\rho\sigma}\mathcal{R}^{\mu\nu\rho\sigma}\nonumber\\
&+c_4\mathcal{R}F_{\mu\nu}F^{\mu\nu}+c_5\mathcal{R}^{\mu\nu}F_{\mu\rho}F_{\nu}^\rho+c_6\mathcal{R}^{\mu\nu\rho\sigma}F_{\mu\nu}F_{\rho\sigma}+c_7(F_{\mu\nu}F^{\mu\nu})^2\nonumber\\
&+c_8(\nabla_\mu F_{\rho\sigma})(\nabla^\mu F^{\rho\sigma})+c_9(\nabla_\mu F_{\rho\sigma})(\nabla^\rho F^{\mu\sigma})),
\end{align}
where the coefficients $c_i$ can depend on the moduli of the underlying theory. Assuming $c_i$'s are stabilized, we can find the correction to the Einstein equations and therefore find $\delta T_c$ as follows.\allowdisplaybreaks
\begin{align}\label{cctt}
    \delta T_c=&-12c_1\Box\mathcal{R}\nonumber\\
    &-4c_2\Box\mathcal{R}\nonumber\\
    &-4c_3\Box\mathcal{R}\nonumber\\
    &-c_4(2\mathcal{R}F^2+6\Box F^2)\nonumber\\&-c_5(2\mathcal{R}^{\alpha\beta}F_{\alpha\rho}\tensor{F}{_\beta^\rho}+2\nabla_\alpha\nabla_\beta(\tensor{F}{^\alpha_\rho}F^{\beta\rho})+\Box F^2)\nonumber\\
    &-2c_6(\mathcal{R}^{\kappa\lambda\rho\sigma}F_{\kappa\lambda}F_{\rho\sigma}+2\nabla_\alpha\nabla_\beta(\tensor{F}{^\alpha_\rho}F^{\beta\rho}))\nonumber\\&+4c_7(F^2)^2\nonumber\\
    &+2c_8((\nabla_\kappa F_{\rho\sigma})(\nabla^\kappa
F^{\rho\sigma})+(\nabla_\kappa F_{\rho\sigma})(\nabla^\rho
F^{\kappa\sigma})-(\nabla^\kappa F_{\kappa\sigma})(\nabla_\rho
F^{\rho\sigma})+F_{\rho\sigma}\Box F^{\rho\sigma})\nonumber\\
&+2c_9((\nabla_\kappa F_{\rho\sigma})(\nabla^\kappa
F^{\rho\sigma})-(\nabla^\kappa F_{\kappa\sigma})(\nabla_\rho
F^{\rho\sigma})+F_{\rho\sigma}\Box F^{\rho\sigma}).
\end{align}

Now we have a good understanding of how the trace of the Einstein field equations can be corrected both classically, and quantum mechanically. In the classical case the correction to all the Einstein equations (and not just the trace) is clear. However, for quantum corrections retrieving the correction to the individual equations of motion is non-trivial and generally unknown. In the next section, we review a technique which allows us to find the correction to the individual equations is some highly symmetric backgrounds.  

\subsection{Retrieving \texorpdfstring{$\expval{T_{\mu\nu}}$}{TEXT} from its trace}\label{23}

As mentioned in subsection \ref{22}, it generally is formidable to find the quantum corrections to all elements of the energy-momentum tensor. Fortunately, it turns out that in some symmetric backgrounds, symmetries are so restricting that allow us to recover the whole energy-momentum tensor from its trace. In this subsection, we will review the method first proposed in \cite{Abedi:2017vtr} to accomplish this task.

We use the symmetry properties of Reissner-Nordstrom to retrieve all the components of $\delta T_{\mu\nu}$ from its trace. Following is a list of expectations that we have for $\delta T^{\mu\nu}$ based on symmetry considerations.
\begin{itemize}
\item \textbf{Static: } As explained in the beginning of the section, we work in adiabatic regime where we can neglect the time evolution of the black hole due to the Hawking radiation and assume it is static. We can engineer such a vacuum by surrounding the black hole with a reflective mirror that traps the Hawking radiation and prevents the black hole from evaporating. This sets $\delta T^{0i}=0$.
\item \textbf{Spherical: } For static spherical backgrounds, $\delta T$ is diagonal and $\delta T^\theta_\theta$ is equal to $\delta T^\phi_\phi$.
\item \textbf{Radial boost invariance: } Reissner-Nordstrom solution is invariant under radial boost. That is to say the Riemann tensor, and $G^{\mu\nu}$, are invariant under radial boost. We assume that this symmetry carries on to the corrected solution so that a radially free-falling observer will not be able to tell their radial velocity based on the vacuum energy-momentum tensor. One can see that this criterion is equivalent to $\delta T^r_r=\delta T^t_t$.
\item \textbf{Conservation:} Finally, $\delta T^{\mu\nu}=\frac{1}{8\pi}G^{\mu\nu}-T^{\mu\nu}$, must be conserved.
\end{itemize}

Using the above properties, one can show that the Taylor expansion of $\delta T^{\mu\nu}$ in terms of $r^{-1}$ has the following structure,
\begin{align}
\label{Tform}
\delta{T}^\mu_\nu=\sum\limits_{\substack{p \geq 0\\p\neq 4}} T_p
\left(\begin{array}{
cc}
\begin{array}{
cc}
I_2
\end{array} & \begin{array}{
cc}
0
\end{array} \\
\begin{array}{
cc}
0
\end{array} & \begin{array}{
cc}
(-\frac{p}{2}+1)I_2
\end{array}
\end{array}\right)
r^{-p}.
\end{align}
Note that we have excluded $p=4$ from the series in the equation \eqref{Tform}, because such a term redefines the electric charge and so should be absorbed in the energy momentum tensor of the electromagnetic field, $T^{\mu\nu}$. All left to find $\expval{T_{\mu\nu}}$ is to find the coefficients $T_P$, which can be read off from the Taylor expansion of the trace as follows,
\begin{align}
\label{traceformula}
\delta{T}^\mu_\mu=\sum\limits_{p \geq 0} (4-p)T_p r^{-p}.
\end{align}
 
In conclusion, we showed how to uniquely find the corrected metric and $\expval{T^{\mu\nu}}$ up to any arbitrary order of perturbation from the expression of $\expval{{T}^\mu_\mu}$.

\section{Corrections to extremal black holes}\label{sec3}
In this section we use the tools developed in the previous section to study the overall impact of different corrections to extremal black holes.

We divide our investigation of corrections to the extremality conditions based on the type of the correction.

\subsubsection{Quantum corrections + renormalized cosmological constant}

In this part, we separately study the first order corrections due to the conformal charges and the cosmological constant. Note that, what we mean by cosmological constant is any curvature-independent contribution to the fields equations, whether quantum or classical. We start with studying the effect of conformal charges.
\vspace{2mm}

\bf a) Conformal charges\normalfont
\vspace{1mm}

After solving the equations for the corrected metric based on the previous subsection, one could find the first order correction to the extremality condition. This correction has been extensively studied in \cite{Abedi:2017vtr} whose results can be expressed in terms of the conformal charges as follows, 

\begin{align}
\label{CCE}
\frac{Q}{M}\simeq 1-\frac{1}{1800\pi}m_P^2\frac{6c-5a}{M^2}.
\end{align}

We complete this subsection by studying the effect of the cosmological constant on the extremality condition. 
\vspace{2mm}

\bf b) Cosmological constant\normalfont
\vspace{1mm}

We will show that for a range of macroscopic black holes, the effect of cosmological constant is negligible compared to that of anomaly coefficients $a$ and $c$. 

Cosmological constant modifies the extremality condition as \cite{Astefanesei:2003gw,Cardoso:2010rz},
\begin{align}
M=\sqrt{\frac{1-\sqrt{1-4{\Lambda Q^2}/{M_P^6}}}{2\Lambda/M_P^6}}(\frac{2+\sqrt{1-4\Lambda Q^2/M_P^6}}{3}).
\end{align}

The above equation points out that there is an upper bound for the charge and mass for stable extremal black holes respectively equal to $\frac{M_P^3}{2}\Lambda^{-\frac{1}{2}}$ and $\frac{\sqrt2M_P^3}{3}\Lambda^{-\frac{1}{2}}$. For masses (and charges) much smaller than this limit, one can safely use perturbation to obtain the first order correction to the extremality condition as follows,
\begin{align}\label{corlam}
\frac{Q}{M}\simeq1+\frac{M^2\Lambda}{6M_P^6}.
\end{align}

The Weak Gravity Conjecture (WGC) ($\frac{Q}{M}>1$) is expected to hold for black holes with masses much greater than the Planck mass. Consider the black holes that have a horizon radius much smaller than $\Lambda^{-1/2}$. Such black holes exist if there is a scale separation between the IR cut-off $\Lambda^{1/2}$ and the UV cutoff $\Lambda_{QG}$ which also sets the inverse radius of the smallest black hole. If such a separation of scales does not exist, the EFT breaks down on cosmological scales, hence we assume that such a separation exists and from this point on, we focus our analysis on black holes that are very small compared to the cosmological horizon but very large compared to the UV scale. The lower bound assures the validity of the WGC, while the upper bound is chosen so that the effect of the cosmological constant given by the equation \eqref{corlam} is heavily dominated by the correction due to the conformal charges given by the equation \eqref{CCE}. Therefore, the effect of the cosmological constant could be neglected in comparison to that of anomaly coefficients.

All in all, our final expression for the correction to $\frac{ Q}{M}$ due to quantum fluctuations and renormalized cosmological constant is

\begin{align}
\label{QuantumC}
\delta(\frac{Q}{M})\simeq-\frac{m_P^2}{1800\pi}\frac{6c-5a}{M^2}.
\end{align}

\subsubsection{Classical corrections to the extremality conditions}

For the general action \eqref{cctt} one can perturbatively solve the effectively corrected Einsetein field equations to find \cite{Kats:2006xp}
\begin{align}\label{ccte}
\frac{Q}{M}\simeq 1+\frac{\delta_c}{M^2},
\end{align}
where 
\begin{align}
    \delta_cl_P^2=\frac{16\pi}{5}c_2+\frac{64\pi}{5}c_3+\frac{2m_P^2}{5}c_5+\frac{2m_P^2}{5}c_6+\frac{m_P^4}{5\pi}c_7-\frac{2m_P^2}{5}c_8-\frac{m_P^2}{5}c_9.
\end{align}
For example, the first order effective action of heterotic theory on $T^6$ is \cite{Gross:1986mw}
\begin{align}
\label{Heterotic}
S^{String}=\int dx^{4} \{&\sqrt {-g}[\frac{1}{16\pi}\mathcal{R}-\frac{1}{4}F_{\mu\nu}F^{\mu\nu}]\nonumber\\
&+\sqrt {-g}h\alpha'[\frac{1}{128\pi}\mathcal{R}^2-\frac{1}{32\pi}\mathcal{R}_{\mu\nu}\mathcal{R}^{\mu\nu}\nonumber\\
&+\frac{m_P^2}{128\pi}\mathcal{R}^{\mu\nu\rho\sigma}\mathcal{R}_{\mu\nu\rho\sigma}+\frac{3\pi}{8m_P^2}(F_{\mu\nu}F^{\mu\nu})^2]\}+...,
\end{align}
 where $h=\exp{-\delta\phi/\sqrt{2}}$ is the rescaled dilaton modulus which we assume is stabilized at $\delta\phi=0$ and $\alpha'=l_s^2$ is the Regge slope. There are many more terms but we have only written the relevant sector of the action. For the action \eqref{Heterotic}, $\delta_c$ is given by,
\begin{align}
\label{ECS}
\delta_c=\frac{6\pi^2h\alpha'}{5l_P^4}\sim \frac{l_s^2}{l_P^4}.
\end{align}

In the next subsection, we will combine the above classical correction to the extremality condition with our previous calculation of the quantum correction. 

\subsubsection{Overall correction to \texorpdfstring{$M=Q$}{TEXT}}

If we combine the results of the previous two subsections we find 
\begin{align}\label{oace}
    \frac{Q}{M}\simeq 1+\frac{\delta_c}{M^2}-\frac{1}{1800\pi}\frac{(6c-5a)}{l_P^2M^2}.
\end{align}
Note that in string theory, as we saw in \eqref{ECS}, the first correction is typically $\sim l_s^2/M^2l_P^4$ while the second correction is $\sim 1/M^2l_P^2$. Thus, the ratio between the two corrections is $(l_s/l_P)^2$. We will come back to this point in \ref{sec5}.

\section{Classical dominance and the Swampland}\label{sec5}

\subsection{Classical Dominance}\label{AC}

In the previous sections we studied different corrections to the gravitational equations of motion. In section \ref{sec2} we explained how the backreaction of vacuum fluctuations could be viewed as corrections to gravitational amplitudes. As we saw, these corrections are purely quantum and much like the running of the couplings, cannot be absorbed in the renormalization of the local effective action. Thus, we can schematically write any amplitude $\mathcal{M}$ with given external momenta $p^\mu_i$ as
\begin{align}
\mathcal{M}(p_i^\mu)\simeq \sum_k (a_{local}^k+a^k_{\beta}\ln(E/\Lambda_{EFT}))f^k_{local}(p_i^\mu)+a^k_{non-local}f^k_{non-local}(p_i^\mu),
\end{align}
where $f^k$ depend on the external momenta such that they scale like $E^k$ at high energies. Moreover, unlike $f_{non-local}^k$, every $f_{local}^k$ has a momentum dependence that can be represented by a local interaction in the Lagrangian. Therefore, the coefficients $a_{local}$ represent the renormalized coefficients of local (potentially higher derivative) gravitational terms in the effective action at some energy scale. However, $a_{non-local}$ and $a_\beta$ are quantum corrections to the equations of motion. In a theory of quantum gravity, the quantum corrections would come from loops while the classical corrections would come from tree-level amplitudes. It is natural to expect that the tree-level amplitudes dominate over the higher-loop amplitudes (see Figure \ref{TA1}), which leads us to propose the following conjecture for any semiclassical theory\footnote{By semiclassical we refer to a theory with a classical gravitational background and quantum fluctuations around that background. The evolution of the gravitational background field is given by classical equation of motion while the evolution of quantum fields is given by a unitary Schrodinger evolution.} consistent with quantum gravity.

\begin{statement3*}
In the background of large extremal black holes, the non-local corrections to the effective gravitational equations of motion are much smaller than the local corrections of the same order in the mass of the black hole.
\end{statement3*}

Classical Dominance implies that the physics in the background of a large extremal balck hole could be well-approximated with classical local field theory. First, let us look at Schwarzschild black holes. Due to the Hawking radiation, these black holes evaporate. However, the evaporation process cannot be captured by any local corrections to the effective equations of motion. This is because the existence of the Hawking radiation depends on the existence of a horizon which is a global property of the spacetime. In fact, there is an easy counterexample to Classical Dominance in Schwarzschild backgrounds in string theory. It is the Heterotic supergravity on $T^6$, where all classical corrections in Schwarzschild background vanish since the higher derivative correction is proportional to the topological Gauss-Bonet term which does not affect the equation of motion. However quantum corrections do not vanish \cite{Abedi:2015yga} in this background as they depend on the Riemann tensor.

One might think that Classical Dominance goes against Weak Gravity Conjecture because Weak Gravity Conjecture states that extremal black holes generically must decay while classical equations say they are stable. However, not only they are not in tension with each other, but they support each other. In quantum gravity, the single particle states (poles of the amplitudes) always have quantized masses and charges. We expect the same to hold for black holes, which are the extension of particles above a certain mass\footnote{See \cite{Bedroya:2022twb} and references therein for more detail.}. Then Weak Gravity Conjecture could be satisfied if there is no black hole that is exactly extremal. Note that the black hole area law, is a statistical statement which says that there are $\sim\exp(A/4)$ states with masses between $M$ and $M+dM$, however, it does not gives us the exact masses. In fact, in the original WGC paper, it was shown that typically in string theory, this is always the case, unless the extremal black hole is allowed to be BPS. We will discuss the connection between the Weak Gravity Conjecture and Classical Dominance in more detail in \ref{51}. 

Let us examine the Classical Dominance more carefully in string theory. As we pointed in the discussion following \eqref{oace}, the ratio of classical to quantum corrections scales like $(l_s/l_P)^2$ which is always much greater than one. In fact, this is true in all even dimensions where trace anomaly exists\footnote{In odd dimensions, the integrand of the radiative correction is an odd function of momentum and therefore vanishes. Another way to see this is that trace anomaly must be a scalar with mass dimension $d$. However, we cannot write down a scalar with odd mass dimension in terms of Riemann tensors.}. To see this, let us start with the string frame action in $D=10$ and compactify to $d$-dimensions. Suppose the internal dimensions have length scale $R$, we find the lower dimensional action scales as
\begin{align}
   S \sim     \frac{R^{D-d}}{l_s^{D-2}g_s^2}\int dx^d \sqrt{G}e^{-2\delta\phi}(\mathcal{R}+\mathcal{O}(\alpha'\mathcal{R}^2)+...)
\end{align}
where $g_s$ is the string coupling and $\delta\phi$ is dilaton scalar field. To go to the Einstein frame we redefine metric as $G_{\mu\nu}\rightarrow e^{-\frac{4\delta\phi}{d-2}}G_{\mu\nu}$ and we find
\begin{align}
       S\sim \frac{R^{D-d}}{l_s^{D-2}g_s^2}\int dx^D \sqrt{G_E}(\mathcal{R_E}+e^{2\delta\phi}\mathcal{O}(\alpha'\mathcal{R}^2)+...)
\end{align}
Therefore, the $d$-dimensional Newton's constant is given by
\begin{align}
    G_N\sim \frac{l_s^{D-2}g_s^2}{R^{D-d}}\sim l_s^{d-2}g_s^2(\frac{l_s}{R})^{D-d},
\end{align}
and the trace anomaly scales like
\begin{align}
    \expval{T}\sim G_N\mathcal{O}(\mathcal{R}^\frac{d}{2})\sim l_s^{d-2}g_s^2(\frac{l_s}{R})^{D-d}\mathcal{O}(\mathcal{R}^\frac{d}{2}).
\end{align}
If we demand the above term to be smaller than the corresponding $\alpha'$ correction, we find
\begin{align}
    l_s^{d-2}g_s^2(\frac{l_s}{R})^{D-d}\ll l_s^{d-2},
\end{align}
which leads to
\begin{align}
    g_s^2(\frac{l_s}{R})^{D-d}\ll 1.
\end{align}
This is always true since $g_s\ll 1$ and $R\gg l_s$. The factor of $g_s^2$ in the relative ratio is an indication of the fact that trace anomaly comes from 1-loop amplitude of strings. The above calculation is a field theory calculation using the string theory EFT. However, the systematics to directly calculate the classical and quantum corrections to equations of motions in string theory is known in many examples and the results satisfy the Classical Dominance \cite{Callan:1986bc,Callan:1988wz,Green:1999pv,Howe:2003cy,Hyakutake:2005rb,Hyakutake:2006aq}. 

Note that in the above argument we did not keep track of the dimensionless constants such as conformal charges. Despite that, the above argument still provides a strong motivation for the Classical Dominance from string theory. As we will see, taking the conformal charges into account leads to a highly non-trivial result, the species bound. 

Even though the Classical Dominance is natural from a quantum gravity perspective, there is no a priori reason from the perspective of low energy field theory for it to be true. This is the defining characteristic of Swampland conjectures as they are supposed to pick out field theories consistent with quantum gravity from the space of otherwise consistent low-energy theories. 

In the above discussion of string theory, we assumed that the higher derivative terms are non-zero in the higher dimensional theory. However, there are situations where because of high level of supersymmetry, the coefficients of higher derivative terms vanish. For example, the coefficients of $\mathcal{O}(\mathcal{R}^2)$ terms in toroidial compactification of type II supergravity ($\mathcal{N}=8$) vanish \cite{Gross:1986iv}. In that case, the Classical Dominance implies that both trace anomaly and running of coupling constants must vanish as well. This is indeed correct, since in theories with $\mathcal{N}=4$, we have $a_\beta=0$ \cite{Seiberg:1988ur} and in supergravity theories with $\mathcal{N}>4$, we have $a=c=0$ \cite{Kallosh:2016xnm}. This provides yet another non-trivial test for the Classical Dominance.

One of the interesting aspects of the Swampland program is the remarkable degree of consistency between the existing web of conjectures. The coherency of any new conjecture with the existing ones serves as an important test. In the rest of this section we study the non-trivial consequences of the Classical Dominance and how they relate to various other Swampland conjectures. In \ref{53} we derive the species bound from the Classical Dominance and in subsection \ref{51}, we show that the mild version of WGC requires the Classical Dominance. Moreover, we show that in case the corrections are sourced by integrating our massive states, Classical Dominance implies the existence of particles that satisfy WGC. 

\subsection{Species bound}\label{53}

As explained in \cite{vandeHeisteeg:2023ubh}, one can estimate the higher-derivative corrections using the quantum gravity cut-off. Suppose the higher derivative terms in the action are generated through a perturbative expansion in some parameter $l_{QG}^2$ in the underlying quantum gravity.
\begin{align}
    S\sim \frac{1}{G_N}\int d^dx \sqrt{G}(\mathcal{R}+l_{QG}^2\mathcal{O}(\mathcal{R}^2)+l_{QG}^4\mathcal{O}(\mathcal{R}^4)+...).
\end{align}
In that case the expansion diverges for curvatures $\mathcal{R}\gtrsim l_{QG}^{-2}$ and the field theory description breaks down. Thus the cutoff of the field theory is given by
\begin{align}\label{SB0}
    \Lambda_{QG}\simeq\frac{1}{l_{QG}}.
\end{align}
On the other hand, the Classical Dominance in 4d implies that 
\begin{align}\label{SB1}
    l_{QG}^2\gtrsim l_{P}^2(6c-5a).
\end{align}
For supersymmetric theories with $N>2$, we have $c-a=0$ \cite{Meissner:2016onk}. This is no coincidence. In fact, one can argue that $c$ and $a$ are generically of the same order \cite{Hofman:2008ar}. We can think of $c$ and $a$ each as as a rough estimate for the number of species. 
\begin{align}\label{aNsp}
    c\sim a\sim N_{species}.
\end{align}
This intuition was used in \cite{vandeHeisteeg:2022btw} to identify a more concrete and computable quantity in 4d $\mathcal{N}=2$ theories. Plugging \eqref{aNsp} back into \eqref{SB1} leads to
\begin{align}
     l_{QG}^2\gtrsim l_{P}^2N_{species}
\end{align}
which when combined with \eqref{SB0} gives
\begin{align}
    \Lambda_{EFT}<\frac{m_P}{\sqrt{N_{species}}}.
\end{align}
This is the species bound \cite{Dvali:2007wp}! To appreciate this coincedence, let us review the original argument behind the species bound which is very different. The bound comes from demanding that a cutoff-scale black hole has entropy larger than the number of species.
\begin{align}\label{SB}
    N_{species}<S(R=\Lambda_{EFT}^{-1})\sim (\frac{m_P}{\Lambda_{EFT}})^2
\end{align}
The Classical Dominance somehow knows about the black hole entropy formula. Surprisingly, the matching between the two bounds extends to higher dimensions. In $d$-dimensions, the trace anomaly (or field theory one loop amplitudes in general) are of the order $G_NN_{species}\sim l_P^{d-2}N_{species}$. The Classical Dominance states that
\begin{align}
    l_P^{d-2}N_{species}<l_{QG}^{d-2}
\end{align}
which leads to 
\begin{align}
    \Lambda<\frac{1}{l_{QG}}<\frac{M_P}{N_{species}^\frac{1}{d-2}}.
\end{align}
This inequality matches with the higher dimensional counterpart of inequality \eqref{SB},
\begin{align}
        N_{species}<S(R=\Lambda_{EFT}^{-1})\sim (\frac{m_P}{\Lambda_{EFT}})^{d-2}.
\end{align}
Interestingly, the Classical Dominance which is formulated in terms of the perturbative expansion of string theory is somehow aware of the black hole entropy formula in higher dimensions as well. 

The above argument shows that the Classical Dominance at its core is a statement about the separation of energy scales in quantum gravity. Phenomenologically, the Classical Dominance suggests existence of beyond standard model physics between at least a few orders of magnitudes below the Planck scale. In a theory of quantum gravity the field theory cutoff cannot go all the way to the Planck scale and the information of this UV higherarchy is embedded in the IR physics (e.g. number of light states).

\subsection{Weak Gravity Conjecture}\label{51}

The weak gravity conjecture (WGC) states that in theories of quantum gravity, there must be a state with charge-to-mass ratio greater than one in Planck units. A version WGC known as the mild version states that extremal black holes satisfy this condition \cite{Aalsma:2020duv,Arkani-Hamed:2021ajd,Hamada:2018dde,Kats:2006xp}. The idea is that the corrections we discussed in \ref{sec3} would  perturb the extremality condition so that $Q>M$ for extremal black holes. This allows us to write the WGC as an inequality in terms of the strength of these corrections. In the following, we show that the mild version of WGC is closely, but non-trivially, related to the Classical Dominance. 

Suppose the mild version of WGC is correct, from \eqref{oace} we find
\begin{align}\label{oawgc}
    \delta_c-\frac{m_P^2}{1800\pi^2}(6c-5a)>0,
\end{align}
where $\delta_c$ is the leading tree level correction and the second term is the 1-loop correction. Suppose the Classical Dominance is incorrect and the second term dominates. Then the mild WGC would require
\begin{align}
6c-5a<0,
\end{align}
Let us focus on supersymmetric theories where the anomaly coefficients are more tractable. Table \ref{table2} shows the value of $6c-5a$ for different multiplets of weakly interacting supersymmetric theories in 4d.
\begin{center}
\small
  \renewcommand{\arraystretch}{1.5}
    \addtolength{\tabcolsep}{1pt}  \captionsetup{width=.9\linewidth}
\begin{longtable}{|c|c|c|c|}\hline \hline
\diaghead{xxxxxxxxxxxxxxxxx}{SUSY}{Multiplets} & \makecell{chiral/hyper\\multiplet} & \makecell{vector\\ multiplet} & \makecell{gravity\\multiplet}\\
\hline\hline
 $\mathcal{N}=1$ &\makecell{$~~~~~~~c=15/2~~~~~~~$\\$a={15/4}$\\$6c-5a={105/4}$} &\makecell{$~~~~~~~c={45/2}~~~~~~~$\\$a={135/4}$\\$6c-5a=-{135/4}$} & \makecell{$~~~~~~~c={1155/2}~~~~~~~$\\$a={1695/4}$\\$6c-5a={5385/4}$}\\\hline
$\mathcal{N}=2$ &\makecell{$c=15$\\$a={15/2}$\\$6c-5a={105/2}$} &\makecell{$c=30$\\$a={75/2}$\\$6c-5a=-{15/2}$} & \makecell{$c=390$\\$a=-{615/2}$\\$6c-5a={1605/2}$}\\\hline
$\mathcal{N}=3$ & &\makecell{$c=45$\\$a=45$\\$6c-5a=45$} & \makecell{$c=225$\\$a=225$\\$6c-5a=225$}\\\hline
$\mathcal{N}=4$ & &\makecell{$c=45$\\$a=45$\\$6c-5a=45$} & \makecell{$c=90$\\$a=90$\\$6c-5a=90$}\\\hline
$\mathcal{N}=5,6$ &  &\makecell{$c=0$\\$a=0$\\$6c-5a=0$} & \makecell{$c=0$\\$a=0$\\$6c-5a=0$}\\\hline
$\mathcal{N}=7,8$ & & & \makecell{$c=0$\\$a=0$\\$6c-5a=0$}\\\hline
 \caption{The conformal charges for massless multiplets. All the multiplets in $\mathcal{N}>2$, have vanishing $c-a=0$ and for $\mathcal{N}>4$, all conformal charges vanish.}
\label{table2}
\end{longtable}
\end{center}
As one can see, $6c-5a$ is always non-negative for theories with $\mathcal{N}>2$. This shows that the mild version of Weak Gravity Conjecture is violated for these theories. As for $\mathcal{N}=2$ theories, we find an interesting inequality in terms of the number of hypermultiplets and vector multiplets. Suppose, the UV completetion is a supergravity with $n_v$ vector multiplets and $n_h$ hypermultiplets, we find

\begin{align}
    107-n_v+7n_h< 0.
\end{align}
However, this inequality is typically violated in $\mathcal{N}=2$ theories. To see why, let us consider two $\mathcal{N}=2$ theories that are related to each other via mirror symmetry. In particular, conisder a type IIA (or B) compactifications on two mirror Calabi-Yau threefolds. Equivalently, we can fix the Calabi-Yau threefold and switching between the type IIA and IIB theories. The "mirroring" maps $(n_h,n_v)$ to $(n_v+1,n_h-1)$ in the 4d theory. Note that these two theories are completely inequivalent theories. Thus, the mild WGC gives for the original theory and its mirror theory read
\begin{align}
    &107-n_v+7n_h<0\nonumber\\
    &113-n_h+7n_v<0.
\end{align}
Suppose they are both correct, we find 
\begin{align}
      &107+7n_h<n_v<\frac{1}{7}(n_h-113),
\end{align}
which is invariant under mirror symmetry $(n_h,n_v)\leftrightarrow(n_v+1,n_h-1)$. However, the above two inequalities imply $862+50n_h<0$ which cannot be true. This shows that the condition $6c-5a<0$ is always violated for at least one of the two mirror theories. Thus, the condition $6c-5a<0$ is generically violated for supersymmetric theories with extended supersymmetry, even at $\mathcal{N}=2$. Thus, if the mild version of WGC is true, the first term in \eqref{oawgc} must dominate which means the Classical Dominance must be correct.

In the above discussion, we showed that the Classical Dominance is closely related to whether the large extremal black holes satisfy WGC (i.e. the mild form of WGC). However, the classical dominance is also closely related to whether there are massive charged particles below the quantum gravity cutoff that satisfy the WGC. In order to see that, let us focus on the contribution of massive charged particles to the conformal charges $a$ and $c$ and compare that to their contribution to higher derivative operators. 

As we explained earlier in the section, integrating out a massive particle below the quantum gravity cut-off has an $\mathcal{O}(1)$ contribution to conformal charges.  However, the contribution of such massive particle to the higher derivative terms $(\mathcal{F}^2)^2$, $(\mathcal{F}\tilde{\mathcal{F}})^2$, and $\mathcal{F}\mathcal{R}$ is respectively proportional to $(q/m)^4$, $(q/m)^4$, and $(q/m)^2$ in Planck units. Therefore, the contribution of local higher-derivative terms is only dominant when there are charged particles that satisfy $q/m\gtrsim1$ in Planck units, which is the statement of WGC for particles. 

\section{Conclusions}\label{sec7}

We proposed the Classical Dominance which states that quantum corrections to the Einstein field equations must be suppressed by the local higher-derivative terms of the same order in the effective action. This is natural in string theory because the higher derivative terms come from tree-level amplitudes while quantum corrections are generated by higher genus surfaces which are suppressed by some power of string coupling. 

In addition to the motivation from string theory, we showed that the Classical Dominance implies the species bound in all even dimensions. Given that the species bound is based on black hole arguments, the Classical Dominance somehow sees the black entropy formula. 

We also showed that violation of the Classical Dominance in four dimensions can lead to a violation of the mild version of the WGC. This is because trace anomalies tend to modify extremal black holes in violation of WGC. Therefore, the mild version of WGC requires the higher derivative corrections to dominate and reverse the effect of trace anomaly. Moreover, if the corrections descend from a UV theory by integrating out massive states, classical dominance implies that some of those particles must satisfy WGC. It would be interesting to find a more fundamental understanding of this connection. It would be interesting to see if the relationship between Classical Dominance and WGC could be directly extended to higher dimensions and higher-form gauge symmetries.

In this work, we only explored the implications of the Classical Dominance for the trace anomaly part of $\mathcal{O}(\mathcal{R}^2)$ corrections. Quantum corrections other than trace anomaly can lead to further non-trivial consequences. It would be interesting to explore the implications of the Classical Dominance for such subleading corrections.

\section*{Acknowledgement}
We thank  Kuroush Allameh, Sera Cremonini, Georges Obied, Zahra Kabiri, Rashmish K. Mishra, Amineh Mohseni, Arad Nasiri, and Max Wiesner for insightful discussions. We are especially thankful to Cumrun Vafa for valuable comments on the draft. 

\bibliographystyle{apsrev4-1}
\bibliography{References} 

\begin{thebibliography}{53}%
\makeatletter
\providecommand \@ifxundefined [1]{%
 \@ifx{#1\undefined}
}%
\providecommand \@ifnum [1]{%
 \ifnum #1\expandafter \@firstoftwo
 \else \expandafter \@secondoftwo
 \fi
}%
\providecommand \@ifx [1]{%
 \ifx #1\expandafter \@firstoftwo
 \else \expandafter \@secondoftwo
 \fi
}%
\providecommand \natexlab [1]{#1}%
\providecommand \enquote  [1]{``#1''}%
\providecommand \bibnamefont  [1]{#1}%
\providecommand \bibfnamefont [1]{#1}%
\providecommand \citenamefont [1]{#1}%
\providecommand \href@noop [0]{\@secondoftwo}%
\providecommand \href [0]{\begingroup \@sanitize@url \@href}%
\providecommand \@href[1]{\@@startlink{#1}\@@href}%
\providecommand \@@href[1]{\endgroup#1\@@endlink}%
\providecommand \@sanitize@url [0]{\catcode `\\12\catcode `\$12\catcode
  `\&12\catcode `\#12\catcode `\^12\catcode `\_12\catcode `\%12\relax}%
\providecommand \@@startlink[1]{}%
\providecommand \@@endlink[0]{}%
\providecommand \url  [0]{\begingroup\@sanitize@url \@url }%
\providecommand \@url [1]{\endgroup\@href {#1}{\urlprefix }}%
\providecommand \urlprefix  [0]{URL }%
\providecommand \Eprint [0]{\href }%
\providecommand \doibase [0]{http://dx.doi.org/}%
\providecommand \selectlanguage [0]{\@gobble}%
\providecommand \bibinfo  [0]{\@secondoftwo}%
\providecommand \bibfield  [0]{\@secondoftwo}%
\providecommand \translation [1]{[#1]}%
\providecommand \BibitemOpen [0]{}%
\providecommand \bibitemStop [0]{}%
\providecommand \bibitemNoStop [0]{.\EOS\space}%
\providecommand \EOS [0]{\spacefactor3000\relax}%
\providecommand \BibitemShut  [1]{\csname bibitem#1\endcsname}%
\let\auto@bib@innerbib\@empty
\bibitem [{\citenamefont {Vafa}(2005)}]{Vafa:2005ui}%
  \BibitemOpen
  \bibfield  {author} {\bibinfo {author} {\bibfnamefont {C.}~\bibnamefont
  {Vafa}},\ }\href@noop {} {\  (\bibinfo {year} {2005})},\ \Eprint
  {http://arxiv.org/abs/hep-th/0509212} {arXiv:hep-th/0509212} \BibitemShut
  {NoStop}%
\bibitem [{\citenamefont {Agmon}\ \emph {et~al.}(2022)\citenamefont {Agmon},
  \citenamefont {Bedroya}, \citenamefont {Kang},\ and\ \citenamefont
  {Vafa}}]{Agmon:2022thq}%
  \BibitemOpen
  \bibfield  {author} {\bibinfo {author} {\bibfnamefont {N.~B.}\ \bibnamefont
  {Agmon}}, \bibinfo {author} {\bibfnamefont {A.}~\bibnamefont {Bedroya}},
  \bibinfo {author} {\bibfnamefont {M.~J.}\ \bibnamefont {Kang}}, \ and\
  \bibinfo {author} {\bibfnamefont {C.}~\bibnamefont {Vafa}},\ }\href@noop {}
  {\  (\bibinfo {year} {2022})},\ \Eprint {http://arxiv.org/abs/2212.06187}
  {arXiv:2212.06187 [hep-th]} \BibitemShut {NoStop}%
\bibitem [{\citenamefont {Bellazzini}\ \emph {et~al.}(2016)\citenamefont
  {Bellazzini}, \citenamefont {Cheung},\ and\ \citenamefont
  {Remmen}}]{Bellazzini:2015cra}%
  \BibitemOpen
  \bibfield  {author} {\bibinfo {author} {\bibfnamefont {B.}~\bibnamefont
  {Bellazzini}}, \bibinfo {author} {\bibfnamefont {C.}~\bibnamefont {Cheung}},
  \ and\ \bibinfo {author} {\bibfnamefont {G.~N.}\ \bibnamefont {Remmen}},\
  }\href {\doibase 10.1103/PhysRevD.93.064076} {\bibfield  {journal} {\bibinfo
  {journal} {Phys. Rev. D}\ }\textbf {\bibinfo {volume} {93}},\ \bibinfo
  {pages} {064076} (\bibinfo {year} {2016})},\ \Eprint
  {http://arxiv.org/abs/1509.00851} {arXiv:1509.00851 [hep-th]} \BibitemShut
  {NoStop}%
\bibitem [{\citenamefont {Adams}\ \emph {et~al.}(2006)\citenamefont {Adams},
  \citenamefont {Arkani-Hamed}, \citenamefont {Dubovsky}, \citenamefont
  {Nicolis},\ and\ \citenamefont {Rattazzi}}]{Adams:2006sv}%
  \BibitemOpen
  \bibfield  {author} {\bibinfo {author} {\bibfnamefont {A.}~\bibnamefont
  {Adams}}, \bibinfo {author} {\bibfnamefont {N.}~\bibnamefont {Arkani-Hamed}},
  \bibinfo {author} {\bibfnamefont {S.}~\bibnamefont {Dubovsky}}, \bibinfo
  {author} {\bibfnamefont {A.}~\bibnamefont {Nicolis}}, \ and\ \bibinfo
  {author} {\bibfnamefont {R.}~\bibnamefont {Rattazzi}},\ }\href {\doibase
  10.1088/1126-6708/2006/10/014} {\bibfield  {journal} {\bibinfo  {journal}
  {JHEP}\ }\textbf {\bibinfo {volume} {10}},\ \bibinfo {pages} {014} (\bibinfo
  {year} {2006})},\ \Eprint {http://arxiv.org/abs/hep-th/0602178}
  {arXiv:hep-th/0602178} \BibitemShut {NoStop}%
\bibitem [{\citenamefont {Bellazzini}\ \emph {et~al.}(2019)\citenamefont
  {Bellazzini}, \citenamefont {Lewandowski},\ and\ \citenamefont
  {Serra}}]{Bellazzini:2019xts}%
  \BibitemOpen
  \bibfield  {author} {\bibinfo {author} {\bibfnamefont {B.}~\bibnamefont
  {Bellazzini}}, \bibinfo {author} {\bibfnamefont {M.}~\bibnamefont
  {Lewandowski}}, \ and\ \bibinfo {author} {\bibfnamefont {J.}~\bibnamefont
  {Serra}},\ }\href {\doibase 10.1103/PhysRevLett.123.251103} {\bibfield
  {journal} {\bibinfo  {journal} {Phys. Rev. Lett.}\ }\textbf {\bibinfo
  {volume} {123}},\ \bibinfo {pages} {251103} (\bibinfo {year} {2019})},\
  \Eprint {http://arxiv.org/abs/1902.03250} {arXiv:1902.03250 [hep-th]}
  \BibitemShut {NoStop}%
\bibitem [{\citenamefont {Hamada}\ \emph {et~al.}(2019)\citenamefont {Hamada},
  \citenamefont {Noumi},\ and\ \citenamefont {Shiu}}]{Hamada:2018dde}%
  \BibitemOpen
  \bibfield  {author} {\bibinfo {author} {\bibfnamefont {Y.}~\bibnamefont
  {Hamada}}, \bibinfo {author} {\bibfnamefont {T.}~\bibnamefont {Noumi}}, \
  and\ \bibinfo {author} {\bibfnamefont {G.}~\bibnamefont {Shiu}},\ }\href
  {\doibase 10.1103/PhysRevLett.123.051601} {\bibfield  {journal} {\bibinfo
  {journal} {Phys. Rev. Lett.}\ }\textbf {\bibinfo {volume} {123}},\ \bibinfo
  {pages} {051601} (\bibinfo {year} {2019})},\ \Eprint
  {http://arxiv.org/abs/1810.03637} {arXiv:1810.03637 [hep-th]} \BibitemShut
  {NoStop}%
\bibitem [{\citenamefont {Arkani-Hamed}\ \emph {et~al.}(2022)\citenamefont
  {Arkani-Hamed}, \citenamefont {Huang}, \citenamefont {Liu},\ and\
  \citenamefont {Remmen}}]{Arkani-Hamed:2021ajd}%
  \BibitemOpen
  \bibfield  {author} {\bibinfo {author} {\bibfnamefont {N.}~\bibnamefont
  {Arkani-Hamed}}, \bibinfo {author} {\bibfnamefont {Y.-t.}\ \bibnamefont
  {Huang}}, \bibinfo {author} {\bibfnamefont {J.-Y.}\ \bibnamefont {Liu}}, \
  and\ \bibinfo {author} {\bibfnamefont {G.~N.}\ \bibnamefont {Remmen}},\
  }\href {\doibase 10.1007/JHEP03(2022)083} {\bibfield  {journal} {\bibinfo
  {journal} {JHEP}\ }\textbf {\bibinfo {volume} {03}},\ \bibinfo {pages} {083}
  (\bibinfo {year} {2022})},\ \Eprint {http://arxiv.org/abs/2109.13937}
  {arXiv:2109.13937 [hep-th]} \BibitemShut {NoStop}%
\bibitem [{\citenamefont {Schwartz}(2014)}]{Schwartz:2014sze}%
  \BibitemOpen
  \bibfield  {author} {\bibinfo {author} {\bibfnamefont {M.~D.}\ \bibnamefont
  {Schwartz}},\ }\href@noop {} {\emph {\bibinfo {title} {{Quantum Field Theory
  and the Standard Model}}}}\ (\bibinfo  {publisher} {Cambridge University
  Press},\ \bibinfo {year} {2014})\BibitemShut {NoStop}%
\bibitem [{\citenamefont {Komargodski}\ and\ \citenamefont
  {Schwimmer}(2011)}]{Komargodski:2011vj}%
  \BibitemOpen
  \bibfield  {author} {\bibinfo {author} {\bibfnamefont {Z.}~\bibnamefont
  {Komargodski}}\ and\ \bibinfo {author} {\bibfnamefont {A.}~\bibnamefont
  {Schwimmer}},\ }\href {\doibase 10.1007/JHEP12(2011)099} {\bibfield
  {journal} {\bibinfo  {journal} {JHEP}\ }\textbf {\bibinfo {volume} {12}},\
  \bibinfo {pages} {099} (\bibinfo {year} {2011})},\ \Eprint
  {http://arxiv.org/abs/1107.3987} {arXiv:1107.3987 [hep-th]} \BibitemShut
  {NoStop}%
\bibitem [{\citenamefont {Han}\ and\ \citenamefont
  {Willenbrock}(2005)}]{Han:2004wt}%
  \BibitemOpen
  \bibfield  {author} {\bibinfo {author} {\bibfnamefont {T.}~\bibnamefont
  {Han}}\ and\ \bibinfo {author} {\bibfnamefont {S.}~\bibnamefont
  {Willenbrock}},\ }\href {\doibase 10.1016/j.physletb.2005.04.040} {\bibfield
  {journal} {\bibinfo  {journal} {Phys. Lett. B}\ }\textbf {\bibinfo {volume}
  {616}},\ \bibinfo {pages} {215} (\bibinfo {year} {2005})},\ \Eprint
  {http://arxiv.org/abs/hep-ph/0404182} {arXiv:hep-ph/0404182} \BibitemShut
  {NoStop}%
\bibitem [{\citenamefont {Dvali}\ and\ \citenamefont
  {Redi}(2008)}]{Dvali:2007wp}%
  \BibitemOpen
  \bibfield  {author} {\bibinfo {author} {\bibfnamefont {G.}~\bibnamefont
  {Dvali}}\ and\ \bibinfo {author} {\bibfnamefont {M.}~\bibnamefont {Redi}},\
  }\href {\doibase 10.1103/PhysRevD.77.045027} {\bibfield  {journal} {\bibinfo
  {journal} {Phys. Rev. D}\ }\textbf {\bibinfo {volume} {77}},\ \bibinfo
  {pages} {045027} (\bibinfo {year} {2008})},\ \Eprint
  {http://arxiv.org/abs/0710.4344} {arXiv:0710.4344 [hep-th]} \BibitemShut
  {NoStop}%
\bibitem [{\citenamefont {Arkani-Hamed}\ \emph {et~al.}(2007)\citenamefont
  {Arkani-Hamed}, \citenamefont {Motl}, \citenamefont {Nicolis},\ and\
  \citenamefont {Vafa}}]{ArkaniHamed:2006dz}%
  \BibitemOpen
  \bibfield  {author} {\bibinfo {author} {\bibfnamefont {N.}~\bibnamefont
  {Arkani-Hamed}}, \bibinfo {author} {\bibfnamefont {L.}~\bibnamefont {Motl}},
  \bibinfo {author} {\bibfnamefont {A.}~\bibnamefont {Nicolis}}, \ and\
  \bibinfo {author} {\bibfnamefont {C.}~\bibnamefont {Vafa}},\ }\href {\doibase
  10.1088/1126-6708/2007/06/060} {\bibfield  {journal} {\bibinfo  {journal}
  {JHEP}\ }\textbf {\bibinfo {volume} {06}},\ \bibinfo {pages} {060} (\bibinfo
  {year} {2007})},\ \Eprint {http://arxiv.org/abs/hep-th/0601001}
  {arXiv:hep-th/0601001} \BibitemShut {NoStop}%
\bibitem [{\citenamefont {Ma}\ \emph {et~al.}(2021)\citenamefont {Ma},
  \citenamefont {Pang},\ and\ \citenamefont {L\"u}}]{Ma:2021opb}%
  \BibitemOpen
  \bibfield  {author} {\bibinfo {author} {\bibfnamefont {L.}~\bibnamefont
  {Ma}}, \bibinfo {author} {\bibfnamefont {Y.}~\bibnamefont {Pang}}, \ and\
  \bibinfo {author} {\bibfnamefont {H.}~\bibnamefont {L\"u}},\ }\href@noop {}
  {\  (\bibinfo {year} {2021})},\ \Eprint {http://arxiv.org/abs/2110.03129}
  {arXiv:2110.03129 [hep-th]} \BibitemShut {NoStop}%
\bibitem [{\citenamefont {Cano}\ \emph {et~al.}(2020)\citenamefont {Cano},
  \citenamefont {Ort\'\i{}n},\ and\ \citenamefont {Ramirez}}]{Cano:2019oma}%
  \BibitemOpen
  \bibfield  {author} {\bibinfo {author} {\bibfnamefont {P.~A.}\ \bibnamefont
  {Cano}}, \bibinfo {author} {\bibfnamefont {T.}~\bibnamefont {Ort\'\i{}n}}, \
  and\ \bibinfo {author} {\bibfnamefont {P.~F.}\ \bibnamefont {Ramirez}},\
  }\href {\doibase 10.1007/JHEP02(2020)175} {\bibfield  {journal} {\bibinfo
  {journal} {JHEP}\ }\textbf {\bibinfo {volume} {02}},\ \bibinfo {pages} {175}
  (\bibinfo {year} {2020})},\ \Eprint {http://arxiv.org/abs/1909.08530}
  {arXiv:1909.08530 [hep-th]} \BibitemShut {NoStop}%
\bibitem [{\citenamefont {Charles}\ and\ \citenamefont
  {Larsen}(2016)}]{Charles:2016wjs}%
  \BibitemOpen
  \bibfield  {author} {\bibinfo {author} {\bibfnamefont {A.~M.}\ \bibnamefont
  {Charles}}\ and\ \bibinfo {author} {\bibfnamefont {F.}~\bibnamefont
  {Larsen}},\ }\href {\doibase 10.1007/JHEP10(2016)142} {\bibfield  {journal}
  {\bibinfo  {journal} {JHEP}\ }\textbf {\bibinfo {volume} {10}},\ \bibinfo
  {pages} {142} (\bibinfo {year} {2016})},\ \Eprint
  {http://arxiv.org/abs/1605.07622} {arXiv:1605.07622 [hep-th]} \BibitemShut
  {NoStop}%
\bibitem [{\citenamefont {Aalsma}(2022)}]{Aalsma:2021qga}%
  \BibitemOpen
  \bibfield  {author} {\bibinfo {author} {\bibfnamefont {L.}~\bibnamefont
  {Aalsma}},\ }\href {\doibase 10.1103/PhysRevD.105.066022} {\bibfield
  {journal} {\bibinfo  {journal} {Phys. Rev. D}\ }\textbf {\bibinfo {volume}
  {105}},\ \bibinfo {pages} {066022} (\bibinfo {year} {2022})},\ \Eprint
  {http://arxiv.org/abs/2111.04201} {arXiv:2111.04201 [hep-th]} \BibitemShut
  {NoStop}%
\bibitem [{\citenamefont {Mirbabayi}(2019)}]{Mirbabayi:2019iae}%
  \BibitemOpen
  \bibfield  {author} {\bibinfo {author} {\bibfnamefont {M.}~\bibnamefont
  {Mirbabayi}},\ }\href@noop {} {\  (\bibinfo {year} {2019})},\ \Eprint
  {http://arxiv.org/abs/1905.02736} {arXiv:1905.02736 [hep-th]} \BibitemShut
  {NoStop}%
\bibitem [{\citenamefont {Aalsma}\ \emph {et~al.}(2021)\citenamefont {Aalsma},
  \citenamefont {Cole}, \citenamefont {Loges},\ and\ \citenamefont
  {Shiu}}]{Aalsma:2020duv}%
  \BibitemOpen
  \bibfield  {author} {\bibinfo {author} {\bibfnamefont {L.}~\bibnamefont
  {Aalsma}}, \bibinfo {author} {\bibfnamefont {A.}~\bibnamefont {Cole}},
  \bibinfo {author} {\bibfnamefont {G.~J.}\ \bibnamefont {Loges}}, \ and\
  \bibinfo {author} {\bibfnamefont {G.}~\bibnamefont {Shiu}},\ }\href {\doibase
  10.1007/JHEP03(2021)085} {\bibfield  {journal} {\bibinfo  {journal} {JHEP}\
  }\textbf {\bibinfo {volume} {03}},\ \bibinfo {pages} {085} (\bibinfo {year}
  {2021})},\ \Eprint {http://arxiv.org/abs/2011.05337} {arXiv:2011.05337
  [hep-th]} \BibitemShut {NoStop}%
\bibitem [{\citenamefont {Lovelock}(1971)}]{Lovelock:1971yv}%
  \BibitemOpen
  \bibfield  {author} {\bibinfo {author} {\bibfnamefont {D.}~\bibnamefont
  {Lovelock}},\ }\href {\doibase 10.1063/1.1665613} {\bibfield  {journal}
  {\bibinfo  {journal} {J. Math. Phys.}\ }\textbf {\bibinfo {volume} {12}},\
  \bibinfo {pages} {498} (\bibinfo {year} {1971})}\BibitemShut {NoStop}%
\bibitem [{\citenamefont {Lovelock}(1972)}]{Lovelock:1972vz}%
  \BibitemOpen
  \bibfield  {author} {\bibinfo {author} {\bibfnamefont {D.}~\bibnamefont
  {Lovelock}},\ }\href {\doibase 10.1063/1.1666069} {\bibfield  {journal}
  {\bibinfo  {journal} {J. Math. Phys.}\ }\textbf {\bibinfo {volume} {13}},\
  \bibinfo {pages} {874} (\bibinfo {year} {1972})}\BibitemShut {NoStop}%
\bibitem [{\citenamefont {Duff}(1977)}]{Duff:1977ay}%
  \BibitemOpen
  \bibfield  {author} {\bibinfo {author} {\bibfnamefont {M.~J.}\ \bibnamefont
  {Duff}},\ }\href {\doibase 10.1016/0550-3213(77)90410-2} {\bibfield
  {journal} {\bibinfo  {journal} {Nucl. Phys. B}\ }\textbf {\bibinfo {volume}
  {125}},\ \bibinfo {pages} {334} (\bibinfo {year} {1977})}\BibitemShut
  {NoStop}%
\bibitem [{\citenamefont {Duff}(1994)}]{Duff:1993wm}%
  \BibitemOpen
  \bibfield  {author} {\bibinfo {author} {\bibfnamefont {M.~J.}\ \bibnamefont
  {Duff}},\ }\href {\doibase 10.1088/0264-9381/11/6/004} {\bibfield  {journal}
  {\bibinfo  {journal} {Class. Quant. Grav.}\ }\textbf {\bibinfo {volume}
  {11}},\ \bibinfo {pages} {1387} (\bibinfo {year} {1994})},\ \Eprint
  {http://arxiv.org/abs/hep-th/9308075} {arXiv:hep-th/9308075} \BibitemShut
  {NoStop}%
\bibitem [{\citenamefont {Wald}(1978)}]{Wald:1978pj}%
  \BibitemOpen
  \bibfield  {author} {\bibinfo {author} {\bibfnamefont {R.~M.}\ \bibnamefont
  {Wald}},\ }\href {\doibase 10.1103/PhysRevD.17.1477} {\bibfield  {journal}
  {\bibinfo  {journal} {Phys. Rev. D}\ }\textbf {\bibinfo {volume} {17}},\
  \bibinfo {pages} {1477} (\bibinfo {year} {1978})}\BibitemShut {NoStop}%
\bibitem [{\citenamefont {Christensen}(1976)}]{Christensen:1976vb}%
  \BibitemOpen
  \bibfield  {author} {\bibinfo {author} {\bibfnamefont {S.~M.}\ \bibnamefont
  {Christensen}},\ }\href {\doibase 10.1103/PhysRevD.14.2490} {\bibfield
  {journal} {\bibinfo  {journal} {Phys. Rev. D}\ }\textbf {\bibinfo {volume}
  {14}},\ \bibinfo {pages} {2490} (\bibinfo {year} {1976})}\BibitemShut
  {NoStop}%
\bibitem [{\citenamefont {Adler}\ \emph {et~al.}(1977)\citenamefont {Adler},
  \citenamefont {Lieberman},\ and\ \citenamefont {Ng}}]{Adler:1976jx}%
  \BibitemOpen
  \bibfield  {author} {\bibinfo {author} {\bibfnamefont {S.~L.}\ \bibnamefont
  {Adler}}, \bibinfo {author} {\bibfnamefont {J.}~\bibnamefont {Lieberman}}, \
  and\ \bibinfo {author} {\bibfnamefont {Y.~J.}\ \bibnamefont {Ng}},\ }\href
  {\doibase 10.1016/0003-4916(77)90313-X} {\bibfield  {journal} {\bibinfo
  {journal} {Annals Phys.}\ }\textbf {\bibinfo {volume} {106}},\ \bibinfo
  {pages} {279} (\bibinfo {year} {1977})}\BibitemShut {NoStop}%
\bibitem [{\citenamefont {Wald}(1977)}]{Wald:1977up}%
  \BibitemOpen
  \bibfield  {author} {\bibinfo {author} {\bibfnamefont {R.~M.}\ \bibnamefont
  {Wald}},\ }\href {\doibase 10.1007/BF01609833} {\bibfield  {journal}
  {\bibinfo  {journal} {Commun. Math. Phys.}\ }\textbf {\bibinfo {volume}
  {54}},\ \bibinfo {pages} {1} (\bibinfo {year} {1977})}\BibitemShut {NoStop}%
\bibitem [{\citenamefont {Deser}\ \emph {et~al.}(1976)\citenamefont {Deser},
  \citenamefont {Duff},\ and\ \citenamefont {Isham}}]{Deser:1976yx}%
  \BibitemOpen
  \bibfield  {author} {\bibinfo {author} {\bibfnamefont {S.}~\bibnamefont
  {Deser}}, \bibinfo {author} {\bibfnamefont {M.~J.}\ \bibnamefont {Duff}}, \
  and\ \bibinfo {author} {\bibfnamefont {C.~J.}\ \bibnamefont {Isham}},\ }\href
  {\doibase 10.1016/0550-3213(76)90480-6} {\bibfield  {journal} {\bibinfo
  {journal} {Nucl. Phys. B}\ }\textbf {\bibinfo {volume} {111}},\ \bibinfo
  {pages} {45} (\bibinfo {year} {1976})}\BibitemShut {NoStop}%
\bibitem [{\citenamefont {Brown}(1977)}]{Brown:1976wc}%
  \BibitemOpen
  \bibfield  {author} {\bibinfo {author} {\bibfnamefont {L.~S.}\ \bibnamefont
  {Brown}},\ }\href {\doibase 10.1103/PhysRevD.15.1469} {\bibfield  {journal}
  {\bibinfo  {journal} {Phys. Rev. D}\ }\textbf {\bibinfo {volume} {15}},\
  \bibinfo {pages} {1469} (\bibinfo {year} {1977})}\BibitemShut {NoStop}%
\bibitem [{\citenamefont {Brown}\ and\ \citenamefont
  {Cassidy}(1977)}]{Brown:1977pq}%
  \BibitemOpen
  \bibfield  {author} {\bibinfo {author} {\bibfnamefont {L.~S.}\ \bibnamefont
  {Brown}}\ and\ \bibinfo {author} {\bibfnamefont {J.~P.}\ \bibnamefont
  {Cassidy}},\ }\href {\doibase 10.1103/PhysRevD.15.2810} {\bibfield  {journal}
  {\bibinfo  {journal} {Phys. Rev. D}\ }\textbf {\bibinfo {volume} {15}},\
  \bibinfo {pages} {2810} (\bibinfo {year} {1977})}\BibitemShut {NoStop}%
\bibitem [{\citenamefont {Capper}\ and\ \citenamefont
  {Duff}(1974)}]{Capper:1974ic}%
  \BibitemOpen
  \bibfield  {author} {\bibinfo {author} {\bibfnamefont {D.~M.}\ \bibnamefont
  {Capper}}\ and\ \bibinfo {author} {\bibfnamefont {M.~J.}\ \bibnamefont
  {Duff}},\ }\href {\doibase 10.1007/BF02748300} {\bibfield  {journal}
  {\bibinfo  {journal} {Nuovo Cim. A}\ }\textbf {\bibinfo {volume} {23}},\
  \bibinfo {pages} {173} (\bibinfo {year} {1974})}\BibitemShut {NoStop}%
\bibitem [{\citenamefont {Pascual}\ \emph {et~al.}(1989)\citenamefont
  {Pascual}, \citenamefont {Taron},\ and\ \citenamefont
  {Tarrach}}]{Pascual:1988ri}%
  \BibitemOpen
  \bibfield  {author} {\bibinfo {author} {\bibfnamefont {P.}~\bibnamefont
  {Pascual}}, \bibinfo {author} {\bibfnamefont {J.}~\bibnamefont {Taron}}, \
  and\ \bibinfo {author} {\bibfnamefont {R.}~\bibnamefont {Tarrach}},\ }\href
  {\doibase 10.1103/PhysRevD.39.2993} {\bibfield  {journal} {\bibinfo
  {journal} {Phys. Rev. D}\ }\textbf {\bibinfo {volume} {39}},\ \bibinfo
  {pages} {2993} (\bibinfo {year} {1989})}\BibitemShut {NoStop}%
\bibitem [{\citenamefont {Christensen}\ and\ \citenamefont
  {Duff}(1979)}]{Christensen:1978md}%
  \BibitemOpen
  \bibfield  {author} {\bibinfo {author} {\bibfnamefont {S.~M.}\ \bibnamefont
  {Christensen}}\ and\ \bibinfo {author} {\bibfnamefont {M.~J.}\ \bibnamefont
  {Duff}},\ }\href {\doibase 10.1016/0550-3213(79)90516-9} {\bibfield
  {journal} {\bibinfo  {journal} {Nucl. Phys. B}\ }\textbf {\bibinfo {volume}
  {154}},\ \bibinfo {pages} {301} (\bibinfo {year} {1979})}\BibitemShut
  {NoStop}%
\bibitem [{\citenamefont {Christensen}\ and\ \citenamefont
  {Duff}(1978)}]{Christensen:1978gi}%
  \BibitemOpen
  \bibfield  {author} {\bibinfo {author} {\bibfnamefont {S.~M.}\ \bibnamefont
  {Christensen}}\ and\ \bibinfo {author} {\bibfnamefont {M.~J.}\ \bibnamefont
  {Duff}},\ }\href {\doibase 10.1016/0370-2693(78)90857-2} {\bibfield
  {journal} {\bibinfo  {journal} {Phys. Lett. B}\ }\textbf {\bibinfo {volume}
  {76}},\ \bibinfo {pages} {571} (\bibinfo {year} {1978})}\BibitemShut
  {NoStop}%
\bibitem [{\citenamefont {Meissner}\ and\ \citenamefont
  {Nicolai}(2017)}]{Meissner:2016onk}%
  \BibitemOpen
  \bibfield  {author} {\bibinfo {author} {\bibfnamefont {K.~A.}\ \bibnamefont
  {Meissner}}\ and\ \bibinfo {author} {\bibfnamefont {H.}~\bibnamefont
  {Nicolai}},\ }\href {\doibase 10.1016/j.physletb.2017.06.031} {\bibfield
  {journal} {\bibinfo  {journal} {Phys. Lett. B}\ }\textbf {\bibinfo {volume}
  {772}},\ \bibinfo {pages} {169} (\bibinfo {year} {2017})},\ \Eprint
  {http://arxiv.org/abs/1607.07312} {arXiv:1607.07312 [gr-qc]} \BibitemShut
  {NoStop}%
\bibitem [{\citenamefont {Kats}\ \emph {et~al.}(2007)\citenamefont {Kats},
  \citenamefont {Motl},\ and\ \citenamefont {Padi}}]{Kats:2006xp}%
  \BibitemOpen
  \bibfield  {author} {\bibinfo {author} {\bibfnamefont {Y.}~\bibnamefont
  {Kats}}, \bibinfo {author} {\bibfnamefont {L.}~\bibnamefont {Motl}}, \ and\
  \bibinfo {author} {\bibfnamefont {M.}~\bibnamefont {Padi}},\ }\href {\doibase
  10.1088/1126-6708/2007/12/068} {\bibfield  {journal} {\bibinfo  {journal}
  {JHEP}\ }\textbf {\bibinfo {volume} {12}},\ \bibinfo {pages} {068} (\bibinfo
  {year} {2007})},\ \Eprint {http://arxiv.org/abs/hep-th/0606100}
  {arXiv:hep-th/0606100} \BibitemShut {NoStop}%
\bibitem [{\citenamefont {Abedi}\ \emph {et~al.}(2017)\citenamefont {Abedi},
  \citenamefont {Arfaei}, \citenamefont {Bedroya}, \citenamefont
  {Mehin-Rasuliani}, \citenamefont {Noorikuhani},\ and\ \citenamefont
  {Salehi-Vaziri}}]{Abedi:2017vtr}%
  \BibitemOpen
  \bibfield  {author} {\bibinfo {author} {\bibfnamefont {J.}~\bibnamefont
  {Abedi}}, \bibinfo {author} {\bibfnamefont {H.}~\bibnamefont {Arfaei}},
  \bibinfo {author} {\bibfnamefont {A.}~\bibnamefont {Bedroya}}, \bibinfo
  {author} {\bibfnamefont {A.}~\bibnamefont {Mehin-Rasuliani}}, \bibinfo
  {author} {\bibfnamefont {M.}~\bibnamefont {Noorikuhani}}, \ and\ \bibinfo
  {author} {\bibfnamefont {K.}~\bibnamefont {Salehi-Vaziri}},\ }\href@noop {}
  {\  (\bibinfo {year} {2017})},\ \Eprint {http://arxiv.org/abs/1707.02545}
  {arXiv:1707.02545 [hep-th]} \BibitemShut {NoStop}%
\bibitem [{\citenamefont {Astefanesei}\ \emph {et~al.}(2004)\citenamefont
  {Astefanesei}, \citenamefont {Mann},\ and\ \citenamefont
  {Radu}}]{Astefanesei:2003gw}%
  \BibitemOpen
  \bibfield  {author} {\bibinfo {author} {\bibfnamefont {D.}~\bibnamefont
  {Astefanesei}}, \bibinfo {author} {\bibfnamefont {R.~B.}\ \bibnamefont
  {Mann}}, \ and\ \bibinfo {author} {\bibfnamefont {E.}~\bibnamefont {Radu}},\
  }\href {\doibase 10.1088/1126-6708/2004/01/029} {\bibfield  {journal}
  {\bibinfo  {journal} {JHEP}\ }\textbf {\bibinfo {volume} {01}},\ \bibinfo
  {pages} {029} (\bibinfo {year} {2004})},\ \Eprint
  {http://arxiv.org/abs/hep-th/0310273} {arXiv:hep-th/0310273} \BibitemShut
  {NoStop}%
\bibitem [{\citenamefont {Cardoso}\ \emph {et~al.}(2009)\citenamefont
  {Cardoso}, \citenamefont {Lemos},\ and\ \citenamefont
  {Marques}}]{Cardoso:2010rz}%
  \BibitemOpen
  \bibfield  {author} {\bibinfo {author} {\bibfnamefont {V.}~\bibnamefont
  {Cardoso}}, \bibinfo {author} {\bibfnamefont {M.}~\bibnamefont {Lemos}}, \
  and\ \bibinfo {author} {\bibfnamefont {M.}~\bibnamefont {Marques}},\ }\href
  {\doibase 10.1103/PhysRevD.80.127502} {\bibfield  {journal} {\bibinfo
  {journal} {Phys. Rev. D}\ }\textbf {\bibinfo {volume} {80}},\ \bibinfo
  {pages} {127502} (\bibinfo {year} {2009})},\ \Eprint
  {http://arxiv.org/abs/1001.0019} {arXiv:1001.0019 [gr-qc]} \BibitemShut
  {NoStop}%
\bibitem [{\citenamefont {Gross}\ and\ \citenamefont
  {Sloan}(1987)}]{Gross:1986mw}%
  \BibitemOpen
  \bibfield  {author} {\bibinfo {author} {\bibfnamefont {D.~J.}\ \bibnamefont
  {Gross}}\ and\ \bibinfo {author} {\bibfnamefont {J.~H.}\ \bibnamefont
  {Sloan}},\ }\href {\doibase 10.1016/0550-3213(87)90465-2} {\bibfield
  {journal} {\bibinfo  {journal} {Nucl. Phys. B}\ }\textbf {\bibinfo {volume}
  {291}},\ \bibinfo {pages} {41} (\bibinfo {year} {1987})}\BibitemShut
  {NoStop}%
\bibitem [{\citenamefont {Abedi}\ and\ \citenamefont
  {Arfaei}(2016)}]{Abedi:2015yga}%
  \BibitemOpen
  \bibfield  {author} {\bibinfo {author} {\bibfnamefont {J.}~\bibnamefont
  {Abedi}}\ and\ \bibinfo {author} {\bibfnamefont {H.}~\bibnamefont {Arfaei}},\
  }\href {\doibase 10.1007/JHEP03(2016)135} {\bibfield  {journal} {\bibinfo
  {journal} {JHEP}\ }\textbf {\bibinfo {volume} {03}},\ \bibinfo {pages} {135}
  (\bibinfo {year} {2016})},\ \Eprint {http://arxiv.org/abs/1506.05844}
  {arXiv:1506.05844 [gr-qc]} \BibitemShut {NoStop}%
\bibitem [{\citenamefont {Bedroya}(2022)}]{Bedroya:2022twb}%
  \BibitemOpen
  \bibfield  {author} {\bibinfo {author} {\bibfnamefont {A.}~\bibnamefont
  {Bedroya}},\ }\href@noop {} {\  (\bibinfo {year} {2022})},\ \Eprint
  {http://arxiv.org/abs/2211.17162} {arXiv:2211.17162 [hep-th]} \BibitemShut
  {NoStop}%
\bibitem [{\citenamefont {Callan}\ \emph {et~al.}(1987)\citenamefont {Callan},
  \citenamefont {Lovelace}, \citenamefont {Nappi},\ and\ \citenamefont
  {Yost}}]{Callan:1986bc}%
  \BibitemOpen
  \bibfield  {author} {\bibinfo {author} {\bibfnamefont {C.~G.}\ \bibnamefont
  {Callan}, \bibfnamefont {Jr.}}, \bibinfo {author} {\bibfnamefont
  {C.}~\bibnamefont {Lovelace}}, \bibinfo {author} {\bibfnamefont {C.~R.}\
  \bibnamefont {Nappi}}, \ and\ \bibinfo {author} {\bibfnamefont {S.~A.}\
  \bibnamefont {Yost}},\ }\href {\doibase 10.1016/0550-3213(87)90227-6}
  {\bibfield  {journal} {\bibinfo  {journal} {Nucl. Phys. B}\ }\textbf
  {\bibinfo {volume} {288}},\ \bibinfo {pages} {525} (\bibinfo {year}
  {1987})}\BibitemShut {NoStop}%
\bibitem [{\citenamefont {Callan}\ \emph {et~al.}(1988)\citenamefont {Callan},
  \citenamefont {Lovelace}, \citenamefont {Nappi},\ and\ \citenamefont
  {Yost}}]{Callan:1988wz}%
  \BibitemOpen
  \bibfield  {author} {\bibinfo {author} {\bibfnamefont {C.~G.}\ \bibnamefont
  {Callan}, \bibfnamefont {Jr.}}, \bibinfo {author} {\bibfnamefont
  {C.}~\bibnamefont {Lovelace}}, \bibinfo {author} {\bibfnamefont {C.~R.}\
  \bibnamefont {Nappi}}, \ and\ \bibinfo {author} {\bibfnamefont {S.~A.}\
  \bibnamefont {Yost}},\ }\href {\doibase 10.1016/0550-3213(88)90565-2}
  {\bibfield  {journal} {\bibinfo  {journal} {Nucl. Phys. B}\ }\textbf
  {\bibinfo {volume} {308}},\ \bibinfo {pages} {221} (\bibinfo {year}
  {1988})}\BibitemShut {NoStop}%
\bibitem [{\citenamefont {Green}\ and\ \citenamefont
  {Vanhove}(2000)}]{Green:1999pv}%
  \BibitemOpen
  \bibfield  {author} {\bibinfo {author} {\bibfnamefont {M.~B.}\ \bibnamefont
  {Green}}\ and\ \bibinfo {author} {\bibfnamefont {P.}~\bibnamefont
  {Vanhove}},\ }\href {\doibase 10.1103/PhysRevD.61.104011} {\bibfield
  {journal} {\bibinfo  {journal} {Phys. Rev. D}\ }\textbf {\bibinfo {volume}
  {61}},\ \bibinfo {pages} {104011} (\bibinfo {year} {2000})},\ \Eprint
  {http://arxiv.org/abs/hep-th/9910056} {arXiv:hep-th/9910056} \BibitemShut
  {NoStop}%
\bibitem [{\citenamefont {Howe}\ and\ \citenamefont
  {Tsimpis}(2003)}]{Howe:2003cy}%
  \BibitemOpen
  \bibfield  {author} {\bibinfo {author} {\bibfnamefont {P.~S.}\ \bibnamefont
  {Howe}}\ and\ \bibinfo {author} {\bibfnamefont {D.}~\bibnamefont {Tsimpis}},\
  }\href {\doibase 10.1088/1126-6708/2003/09/038} {\bibfield  {journal}
  {\bibinfo  {journal} {JHEP}\ }\textbf {\bibinfo {volume} {09}},\ \bibinfo
  {pages} {038} (\bibinfo {year} {2003})},\ \Eprint
  {http://arxiv.org/abs/hep-th/0305129} {arXiv:hep-th/0305129} \BibitemShut
  {NoStop}%
\bibitem [{\citenamefont {Hyakutake}\ and\ \citenamefont
  {Ogushi}(2006{\natexlab{a}})}]{Hyakutake:2005rb}%
  \BibitemOpen
  \bibfield  {author} {\bibinfo {author} {\bibfnamefont {Y.}~\bibnamefont
  {Hyakutake}}\ and\ \bibinfo {author} {\bibfnamefont {S.}~\bibnamefont
  {Ogushi}},\ }\href {\doibase 10.1103/PhysRevD.74.025022} {\bibfield
  {journal} {\bibinfo  {journal} {Phys. Rev. D}\ }\textbf {\bibinfo {volume}
  {74}},\ \bibinfo {pages} {025022} (\bibinfo {year} {2006}{\natexlab{a}})},\
  \Eprint {http://arxiv.org/abs/hep-th/0508204} {arXiv:hep-th/0508204}
  \BibitemShut {NoStop}%
\bibitem [{\citenamefont {Hyakutake}\ and\ \citenamefont
  {Ogushi}(2006{\natexlab{b}})}]{Hyakutake:2006aq}%
  \BibitemOpen
  \bibfield  {author} {\bibinfo {author} {\bibfnamefont {Y.}~\bibnamefont
  {Hyakutake}}\ and\ \bibinfo {author} {\bibfnamefont {S.}~\bibnamefont
  {Ogushi}},\ }\href {\doibase 10.1088/1126-6708/2006/02/068} {\bibfield
  {journal} {\bibinfo  {journal} {JHEP}\ }\textbf {\bibinfo {volume} {02}},\
  \bibinfo {pages} {068} (\bibinfo {year} {2006}{\natexlab{b}})},\ \Eprint
  {http://arxiv.org/abs/hep-th/0601092} {arXiv:hep-th/0601092} \BibitemShut
  {NoStop}%
\bibitem [{\citenamefont {Gross}\ and\ \citenamefont
  {Witten}(1986)}]{Gross:1986iv}%
  \BibitemOpen
  \bibfield  {author} {\bibinfo {author} {\bibfnamefont {D.~J.}\ \bibnamefont
  {Gross}}\ and\ \bibinfo {author} {\bibfnamefont {E.}~\bibnamefont {Witten}},\
  }\href {\doibase 10.1016/0550-3213(86)90429-3} {\bibfield  {journal}
  {\bibinfo  {journal} {Nucl. Phys. B}\ }\textbf {\bibinfo {volume} {277}},\
  \bibinfo {pages} {1} (\bibinfo {year} {1986})}\BibitemShut {NoStop}%
\bibitem [{\citenamefont {Seiberg}(1988)}]{Seiberg:1988ur}%
  \BibitemOpen
  \bibfield  {author} {\bibinfo {author} {\bibfnamefont {N.}~\bibnamefont
  {Seiberg}},\ }\href {\doibase 10.1016/0370-2693(88)91265-8} {\bibfield
  {journal} {\bibinfo  {journal} {Phys. Lett. B}\ }\textbf {\bibinfo {volume}
  {206}},\ \bibinfo {pages} {75} (\bibinfo {year} {1988})}\BibitemShut
  {NoStop}%
\bibitem [{\citenamefont {Kallosh}(2017)}]{Kallosh:2016xnm}%
  \BibitemOpen
  \bibfield  {author} {\bibinfo {author} {\bibfnamefont {R.}~\bibnamefont
  {Kallosh}},\ }\href {\doibase 10.1103/PhysRevD.95.041701} {\bibfield
  {journal} {\bibinfo  {journal} {Phys. Rev. D}\ }\textbf {\bibinfo {volume}
  {95}},\ \bibinfo {pages} {041701} (\bibinfo {year} {2017})},\ \Eprint
  {http://arxiv.org/abs/1612.08978} {arXiv:1612.08978 [hep-th]} \BibitemShut
  {NoStop}%
\bibitem [{\citenamefont {van~de Heisteeg}\ \emph {et~al.}(2023)\citenamefont
  {van~de Heisteeg}, \citenamefont {Vafa},\ and\ \citenamefont
  {Wiesner}}]{vandeHeisteeg:2023ubh}%
  \BibitemOpen
  \bibfield  {author} {\bibinfo {author} {\bibfnamefont {D.}~\bibnamefont
  {van~de Heisteeg}}, \bibinfo {author} {\bibfnamefont {C.}~\bibnamefont
  {Vafa}}, \ and\ \bibinfo {author} {\bibfnamefont {M.}~\bibnamefont
  {Wiesner}},\ }\href@noop {} {\  (\bibinfo {year} {2023})},\ \Eprint
  {http://arxiv.org/abs/2303.13580} {arXiv:2303.13580 [hep-th]} \BibitemShut
  {NoStop}%
\bibitem [{\citenamefont {Hofman}\ and\ \citenamefont
  {Maldacena}(2008)}]{Hofman:2008ar}%
  \BibitemOpen
  \bibfield  {author} {\bibinfo {author} {\bibfnamefont {D.~M.}\ \bibnamefont
  {Hofman}}\ and\ \bibinfo {author} {\bibfnamefont {J.}~\bibnamefont
  {Maldacena}},\ }\href {\doibase 10.1088/1126-6708/2008/05/012} {\bibfield
  {journal} {\bibinfo  {journal} {JHEP}\ }\textbf {\bibinfo {volume} {05}},\
  \bibinfo {pages} {012} (\bibinfo {year} {2008})},\ \Eprint
  {http://arxiv.org/abs/0803.1467} {arXiv:0803.1467 [hep-th]} \BibitemShut
  {NoStop}%
\bibitem [{\citenamefont {van~de Heisteeg}\ \emph {et~al.}(2022)\citenamefont
  {van~de Heisteeg}, \citenamefont {Vafa}, \citenamefont {Wiesner},\ and\
  \citenamefont {Wu}}]{vandeHeisteeg:2022btw}%
  \BibitemOpen
  \bibfield  {author} {\bibinfo {author} {\bibfnamefont {D.}~\bibnamefont
  {van~de Heisteeg}}, \bibinfo {author} {\bibfnamefont {C.}~\bibnamefont
  {Vafa}}, \bibinfo {author} {\bibfnamefont {M.}~\bibnamefont {Wiesner}}, \
  and\ \bibinfo {author} {\bibfnamefont {D.~H.}\ \bibnamefont {Wu}},\
  }\href@noop {} {\  (\bibinfo {year} {2022})},\ \Eprint
  {http://arxiv.org/abs/2212.06841} {arXiv:2212.06841 [hep-th]} \BibitemShut
  {NoStop}%
\end{thebibliography}%
\end{document}